\documentclass[11pt,letter]{article}
\usepackage[numbers,sort&compress]{natbib}
\usepackage{multirow}
\usepackage{amssymb}
\usepackage{amsmath,amsfonts}
\usepackage{url}
\usepackage{longtable}
\usepackage{epsfig}
\usepackage{graphicx,colordvi}
\usepackage[dvips]{color}
\usepackage{bm}
\usepackage{pdflscape}
\usepackage{multirow}

\def\t{\mbox{tr}}

\def\cG0{{\cal G}_0}

\def\cG{{\cal G}}

\def \sriro{Sr${}_2$IrO$_{4}$}
\def \srrho{Sr${}_2$RhO$_{4}$}
\def \srruo{Sr${}_2$RuO$_{4}$}
\def \t2g{t$_{2g}$}
\def \eg{e$_{g}$}
\def \dxy{d$_{xy}$}
\def \dxz{d$_{xz}$}
\def \dyz{d$_{yz}$}
\def \dx2y2{d$_{x^2-y^2}$}

\def \j32{$j_{\textrm{eff}}$=$3/2$}
\def \jeff12{$j_{\textrm{eff}}$=$1/2$}

\begin{document}
\parskip 1ex
\pagestyle{empty}

\begin{center}
\includegraphics[clip=true,scale=0.15,angle=0]{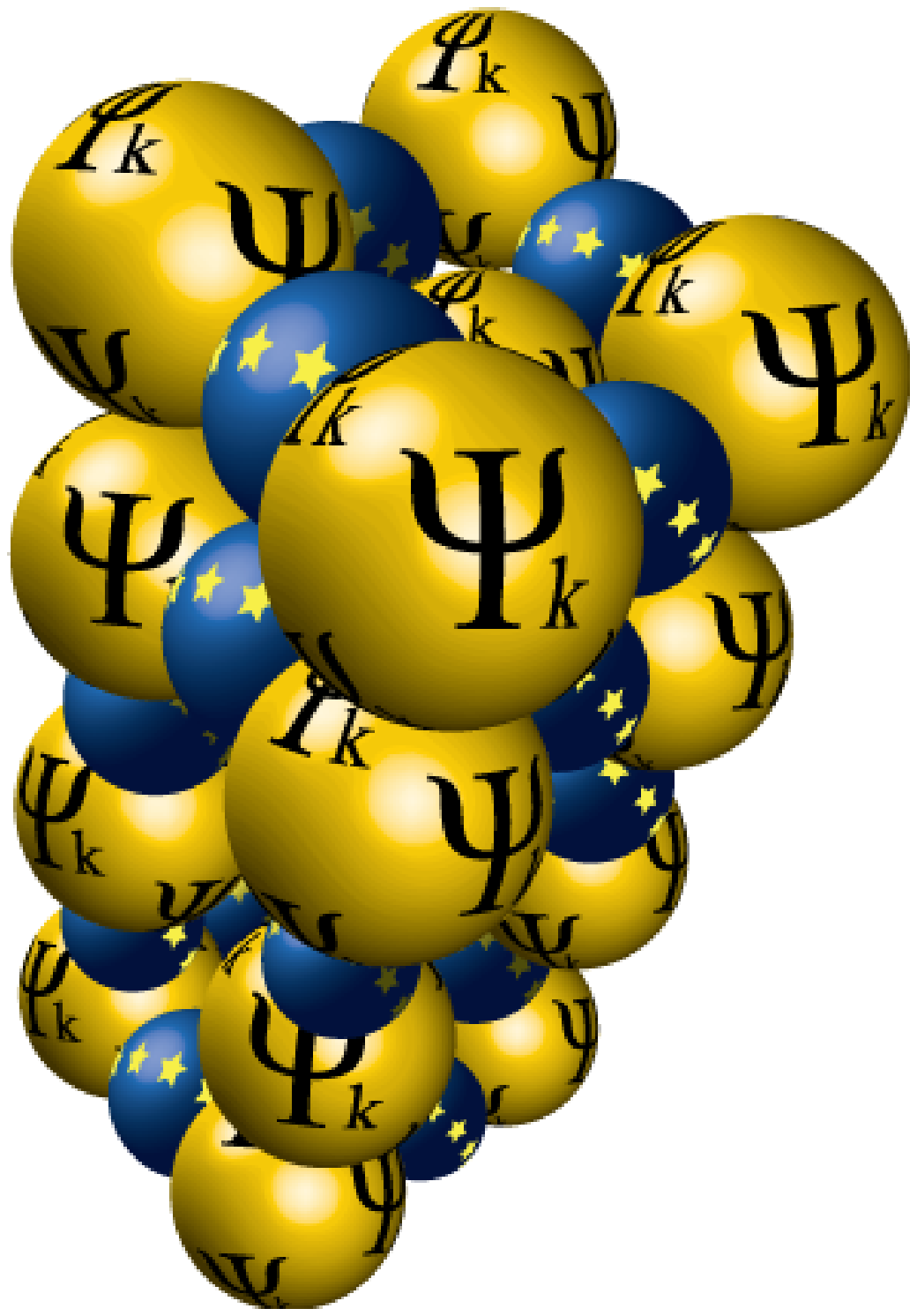}
\hspace{0cm}
\bf \LARGE{$\Psi_k$ Scientific Highlight Of The Month}
\end{center}

\subsection*{\footnotesize No. 132 \hfill November 2016}
\vspace{-0.7cm}
\rule{16.0cm}{0.05mm}
\vspace{1mm}
\setcounter{section}{0}
\setcounter{figure}{0}
\newcommand{\D}{\ensuremath\mathrm{d}}
\begin{center}
{\large \bf Coulomb Correlations in 4d and 5d Oxides from First Principles - or How Spin-Orbit Materials choose their Effective Orbital Degeneracies}

{Cyril Martins$^{1,}$, Markus Aichhorn$^{2}$, and Silke Biermann$^{3,}\footnote{silke.biermann@polytechnique.edu}$}

{\it\small $^1$Laboratoire de Chimie et Physique Quantiques, UMR 5626, Universit\'{e} Paul Sabatier, 118 route de Narbonne, 31400 Toulouse, France}\\
{\it\small $^2$Institute of Theoretical and Computational Physics, Technical University Graz, Petersgasse 16, Graz, Austria}\\
{\it\small $^3$Centre de Physique Th\'{e}orique, Ecole Polytechnique, CNRS UMR 7644, Universit\'{e} Paris-Saclay, 91128 Palaiseau, France}
\end{center}

\begin{abstract}
The interplay of spin-orbit interactions and Coulomb correlations has become a hot topic
in condensed matter theory. Here, we review recent advances in dynamical mean-field theory-based
electronic structure calculations for iridates and rhodates. We stress the notion of
the {\it effective degeneracy} of the compounds, which introduces an additional axis into the
conventional picture of a phase diagram based on filling and on the ratio of interactions to
bandwidth.
\end{abstract}


\section{Introduction}

\begin{figure}
 \small
 \begin{center}
  \includegraphics[width=0.5\linewidth,keepaspectratio]{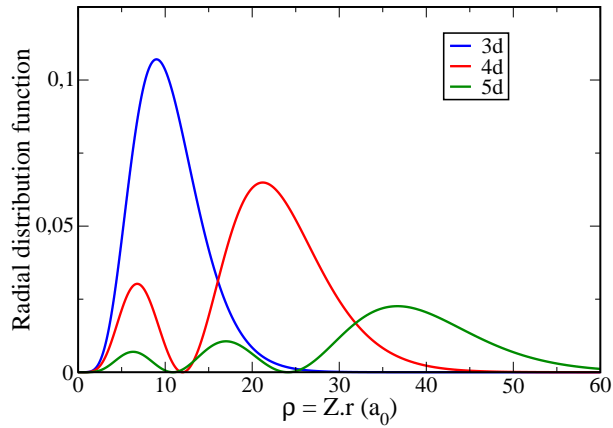}
 \end{center}
 \caption{ \label{fig0} \small Radial distribution function $r^2R_{n\ell}(r)^2$ as a function of the distance from the nucleus $r$ expressed in atomic units, for the $3d$, $4d$ and $5d$ orbitals. To ease the comparison between different atoms, we use the renormalized distance $\rho=Z.r$ on the abscissa, where $Z$ is the effective nuclear charge for a given multi-electron atom. As the principal quantum number $n$ increases, $Z$ remains almost constant for $d$ valence electrons and their radial distribution is thus more and more extended.
} 
\end{figure}

Electronic Coulomb correlations are at the heart of a variety
of exotic properties in compounds with partially filled 3d or 4f
shells.
Prominent examples are found among the 3d transition metal oxides,
where unconventional transport behaviors, ordering phenomena
or unusual spectroscopic properties are observed \cite{ImadaRMP70-98}.
It was argued early on that the comparably weak spatial extension
of 3d orbitals leads to large electronic Coulomb interactions,
competing with kinetic contributions.
Depending on crystal fields, hybridisation, Hund's exchange,
and band filling, this interplay can lead to renormalised metallic 
behavior such as in simple oxides like SrVO$_3$ \cite{TomczakEPL100-12, TomczakPRB90-14} 
or iron pnictide compounds \cite{AichhornPRB82-10, PhysRevLett.113.266403, 
PhysRevB.93.245139, PhysRevB.91.214502, PhysRevLett.113.266407, haule11} 
or induce Mott insulating behavior 
like in YTiO$_3$ \cite{PavariniPRL92-04} or
V$_2$O$_3$ \cite{PoteryaevPRB76-07, arpes_review,keller:205116,PhysRevLett.86.5345}.
According to common belief until recently, such effects would be
less dramatic in 4d and even less in 5d compounds, due to the
substantially more extended radial wave functions of those shells, as shown in Fig.~\ref{fig0}.
The discovery of Mott insulating behavior in Sr$_2$IrO$_4$
therefore triggered a little revolution in the field 
\cite{KimScience323-09,KimPRL101-08}.
In 5d oxides, spin-orbit coupling acts on an energy scale
comparable to the other scales of the system (Coulomb interactions,
bandwidths, ligand fields ...), and the electronic state is the 
result of a complex interplay of Coulomb correlations, spin-orbit
splitting and crystal field effects (for recent reviews see
\cite{Rau, Witzak}).
But, as pointed out already earlier \cite{HaverkortPRL101-08,LiuPRL101-08}, 
also in 4d compounds spin-orbit interactions can influence
the electronic properties substantially. 
In Sr$_2$RhO$_4$, for example, the experimentally observed Fermi
surface can only be reconciled with experiments when spin-orbit
coupling and electronic Coulomb correlations are taken into account
\cite{HaverkortPRL101-08,LiuPRL101-08,MartinsPRL107-11,AhnKunesJPCM27-15}.
Here, we give a review of recent efforts to describe correlated 
spin-orbit physics from first principles, in a combined density functional and
dynamical mean-field theory framework \cite{MartinsPRL107-11}.

\section{Spin-orbit materials -- an incomplete literature review}

The term {\it spin-orbit material} refers to systems 
where spin-orbit coupling (SOC) and its interplay with other elements
of the electronic structure -- crystal or ligand fields, Coulomb
correlations, magnetism, ... -- is essential in determining the
physical properties. 
In many such materials, the physics is largely determined by the
geometrical aspects of the cristalline structure, and the electronic 
properties can be understood by analysing the one-particle band 
structure. In particular, strong enough spin-orbit coupling can 
cause band inversions, possibly leading to
non-trivial topological effects. The quest for topological materials
is nowadays a hot topic of condensed matter physics, and several 
excellent reviews exist in the literature \cite{Ando,RevModPhys.82.3045,
RevModPhys.83.1057}.

The scope of the present review is however a different one. Here, we focus
on materials, where the interplay of spin-orbit interactions
and Coulomb correlations is crucial, and the band picture is at
best useful as a starting point for further many-body calculations.
Early examples are found among the layered tantalum chalcogenides:
TaS$_2$ \cite{PhysRevLett.97.067402, Ritschel, Ma-TaS2}
is Mott insulating thanks to the presence of a lone narrow band resulting
from the combined effect of SOC and a charge-density wave instability.
The corresponding selenide, TaSe$_2$ \cite{PhysRevLett.90.166401}
displays a surface Mott metal-insulator transition.
Nevertheless, the true power of the interplay of spin-orbit 
interactions was fully appreciated only after 
the discovery of \sriro: the insulating behavior -- despite of
moderate Coulomb interactions usually present in 5d compounds --
was even more intriguing, as the electronic and crystal structures
are otherwise seemingly simple.
The interplay of Coulomb correlations and spin-orbit coupling 
was indeed shown to be essential to drive the system insulating,
leading to a state dubbed ``spin-orbit Mott insulator''
\cite{KimScience323-09,KimPRL101-08}. 
A flury of further spin-orbit materials have by now been
characterized, or known compounds have been reinvestigated
in the light of new insights. Iridium-based materials, where
several families of compounds have been studied systematically, 
still hold a privileged position. Tab.~\ref{tab-classIr} summarizes the 
structural, transport and magnetic properties of a selection of 
iridates. It is interesting to note that the large majority among
them display insulating phases. The Ir$^{4+}$ (5d$^5$) state does 
not allow for a band insulating state without symmetry breaking,
and magnetic order is an obvious candidate for helping in opening
the gap. Nevertheless, few compounds have been unambigously
characterized as Slater insulators. 

Slightly more recently, attention focussed yet onto another class 
of 5d materials, namely osmium-based compounds. In this class
fall for example ferroelectric LiOsO$_3$ \cite{Shi-NMat-13} 
as well as the prototypical Slater insulator NaOsO$_3$ 
\cite{PhysRevB.80.161104,JungPRB87-13,CalderPRL108-12,DuPRB85-12,
VecchioSciR3-13,CalderNatComm6-15}
where the loss of magnetic order with increasing temperature is 
accompanied by a closure of the insulating gap.
It has been realised, however, that SOC can also have notable
effects in 4d compounds, with prominent examples among ruthenium-
and rhodium-based materials, where most interesting consequences
for magnetic excitations have been discussed \cite{PhysRevLett.111.197201}.
Tab.~\ref{tab-classOs} gives an overview of the properties of a selection
of osmates, ruthanates and rhodates.
In the following discussion, we will restrict ourselves to the
prototypical correlated iridate \sriro\ and its 4d analog, \srrho.

\begin{table}[!ht]
  \tiny
  \begin{center}
   \begin{tabular}{|c|c c|l l|l l|c|}
    \hline
    \hline
    \multicolumn{8}{|c|}{\multirow{2}{*}{Iridium-based spin-orbit materials}} \\ 
    \multicolumn{8}{|c|}{} \\
    \hline
    Compound & \multicolumn{2}{|c|}{Crystal Struct.} & \multicolumn{2}{|c|}{Transport Property} & \multicolumn{2}{|c|}{Magnetic Ordering} & Ref. \\
\hline
\hline
    CaIrO$_3$ & post-perovskite & $Cmcm$ & Ins. & gap: $0.34$\,eV & AFM & $T_N=115$\,K & \cite{SubediPRB85-12,MorettiSalaPRL112-14,KimPRL115-15} \\
    NaIrO$_3$ & post-perovskite & $Cmcm$ & Ins. & -- & \multicolumn{2}{|c|}{None}& \cite{BremholmJSSC3-11,DuEPL101-13} \\
    BaIrO$_3$ & monoclinic & $C2/m$ & Ins. & gap: $0.05$\,eV & FM & $T_C=180$\,K & \cite{MaitiPRL95-05,ChengPRB80-09,JuPRB87-13} \\
    SrIrO$_3$ & monoclinic & $C2/c$ & Metal & & \multicolumn{2}{|c|}{None}&  \cite{MoonPRL101-08,CaoPRB76-07,NiePRL114-15,MoonJKPS64-14} \\ 
\hline
    $\alpha$-Na$_2$IrO$_3$ & honeycomb monoclinic & $C2/c$ & Ins. & gap: $0.35$\,eV & zig-zag AFM &  $T_N=15$\,K & \cite{SinghPRB82-10,CominPRL109-12,LiuPRB83-11,GretarsssonPRL110-13,
PhysRevB.92.024413,PhysRevB.94.064435}\\
    $\alpha$-Li$_2$IrO$_3$ & honeycomb monoclinic & $C2/c$ & Ins. & -- & spiral AFM & $T_N=15$\,K &  \cite{SinghPRL108-12,ReutherPRB90-14}\\
    $\beta$-Li$_2$IrO$_3$ & hyperhoneycomb & $Fddd$ & Ins. & -- & unconventional AFM & $T_N=38$\,K & \cite{BiffinPRB90-14,TakayamaPRL114-15} \\
    $\gamma$-Li$_2$IrO$_3$ & stripyhoneycomb & $Cccm$ & Ins. & -- & unconventional AFM & $T_N=38$\,K & \cite{BiffinPRL113-14} \\
\hline
    Ba$_2$IrO$_4$ & K$_2$NiF$_4$-type & $I4/mmm$ & Ins. & gap: $0.14$\,eV & AFM & $T_N=240$\,K &  \cite{OkabePRB84-11,AritaPRL108-12,KatukuriPRB85-12,MoserNJP16-14} \\
    Sr$_2$IrO$_4$ & distorted K$_2$NiF$_4$-type & $I4_1/acd$ & Ins. & gap $0.25$\,eV & canted AFM & $T_N=240$\,K  & \cite{KimPRL101-08,KimScience323-09,MartinsPRL107-11} \\
\hline
    Ca$_4$IrO$_6$ & hexagonal & $R\bar{3}c$ & Ins. & -- & AFM & $T_N=13.95$\,K & \cite{CalderPRB89-14,CaoPRB75-07,FrankePRB83-11} \\ 
\hline
    Y$_2$Ir$_2$O$_7$ & pyrochlore & $Fd\bar{3}m$ & Ins. & -- & AIAO & $T=155$\,K & \cite{ShapiroPRB85-12,DisselerPRB89-14} \\
    Pr$_2$Ir$_2$O$_7$ & pyrochlore & $Fd\bar{3}m$ & Metal &  & \multicolumn{2}{|c|}{None}& \cite{MatsuhiraJPSJ80-11,KondoNComm6-15} \\
    Nd$_2$Ir$_2$O$_7$ & pyrochlore & $Fd\bar{3}m$ & MIT & $T=36$\,K & AIAO & $T=36$\,K & \cite{MatsuhiraJPSJ76-07,MatsuhiraJPSJ80-11} \\
    Sm$_2$Ir$_2$O$_7$ & pyrochlore & $Fd\bar{3}m$ & MIT & $T=117$\,K & AIAO & $T=117$\,K & \cite{MatsuhiraJPSJ76-07,MatsuhiraJPSJ80-11} \\
    Eu$_2$Ir$_2$O$_7$ & pyrochlore & $Fd\bar{3}m$ & MIT & $T=120$\,K & AIAO & $T=120$\,K & \cite{MatsuhiraJPSJ80-11,SagayamaPRB87-13,SushkovPRB92-15,UematsuPRB92-15}\\
    Gd$_2$Ir$_2$O$_7$ & pyrochlore & $Fd\bar{3}m$ & Ins. & -- & AIAO & $T=127$\,K & \cite{MatsuhiraJPSJ80-11} \\
    Tb$_2$Ir$_2$O$_7$ & pyrochlore & $Fd\bar{3}m$ & Ins. & -- & AIAO & $T=130$\,K & \cite{MatsuhiraJPSJ80-11,LeFrancoisPRL114-15} \\
    Dy$_2$Ir$_2$O$_7$ & pyrochlore & $Fd\bar{3}m$ & Ins. & -- & AIAO & $T=134$\,K & \cite{MatsuhiraJPSJ80-11} \\
    Ho$_2$Ir$_2$O$_7$ & pyrochlore & $Fd\bar{3}m$ & Ins. & -- & AIAO & $T=141$\,K & \cite{MatsuhiraJPSJ80-11} \\
    Er$_2$Ir$_2$O$_7$ & pyrochlore & $Fd\bar{3}m$ & Ins. & -- & AIAO & $T=140$\,K & \cite{LeFrancoisPRL114-15} \\
    Yb$_2$Ir$_2$O$_7$ & pyrochlore & $Fd\bar{3}m$ & Ins. & -- & AIAO & $T=130$\,K & \cite{DisselerPRB86-12} \\
    Lu$_2$Ir$_2$O$_7$ & pyrochlore & $Fd\bar{3}m$ & Ins. & -- & AIAO & $T=120$\,K & \cite{TairaJPCM23-01} \\
    Bi$_2$Ir$_2$O$_7$ & pyrochlore & $Fd\bar{3}m$ & Metal & & \multicolumn{2}{|c|}{None}& \cite{LeePRB87-13-2,QiJPCM14-12} \\
\hline
    Sr$_3$Ir$_2$O$_7$ & monoclinic & $C2/c$ & Ins. & gap: $0.1$\,eV & AFM & $T_N=285$\,K & \cite{CaoPRB66-02,MoonPRL101-08,FujiyamaPRB86-12,ParkPRB89-14,HoganPRB93-16,
PhysRevLett.109.157402} \\
\hline
    Na$_4$Ir$_3$O$_8$ & hyperkagome & $P4_132$ & Ins. & -- & AFM & $T_N=6$\,K &  \cite{OkamotoPRL99-07,DallyPRL113-14,BalodhiPRB91-15} \\    
\hline
    Ca$_5$Ir$_3$O$_{12}$ & hexagonal & $P-62m$ & Ins. & -- & AFM & $T_N=7.8$\,K & \cite{WakeshimaSSC125-03,CaoPRB75-07,FrankePRB83-11} \\
\hline
    La$_2$ZnIrO$_6$ & double-perovskite & $P2_1/n$ & Ins. & -- & FM & $T_C=7.5$\,K & \cite{CaoPRB87-13} \\
    La$_2$MgIrO$_6$ & double-perovskite & $P2_1/n$ & Ins. & gap: $0.16$\,eV & AFM & $T_N=12$\,K & \cite{CaoPRB87-13,Ghimire-PRB93-16} \\
    Pr$_2$MgIrO$_6$ & double-perovskite & $P2_1/n$ & Ins. & gap: $0.2$\,eV & AFM & $T_N=14$\,K & \cite{MugaveroIII465-10,Ghimire-PRB93-16} \\
    Nd$_2$MgIrO$_6$ & double-perovskite & $P2_1/n$ & Ins. & -- & AFM & $T_N=12$\,K & \cite{MugaveroIII465-10} \\
    Sm$_2$MgIrO$_6$ & double-perovskite & $P2_1/n$ & Ins. & --  & AFM & $T_N=15$\,K & \cite{MugaveroIII465-10} \\
    Eu$_2$MgIrO$_6$ & double-perovskite & $P2_1/n$ & Ins. & --  & AFM & $T_N=10$\,K & \cite{MugaveroIII465-10} \\
    Gd$_2$MgIrO$_6$ & double-perovskite & $P2_1/n$ & Ins. & --  & \multicolumn{2}{|c|}{None}& \cite{MugaveroIII465-10} \\
    Sr$_2$CeIrO$_6$ & double perovskite & $P2_1/n$ & Ins. & gap: $0.3$\,eV & AFM & $T_N=21$\,K & \cite{HaradaJSSC145-99,PandaMPLB27-13,KanungoArXiv-15} \\
    Ba$_2$YIrO$_6$ & double perovskite & $Fm\bar{3}m$ & Ins. & gap: $0.221$\,eV & \multicolumn{2}{|c|}{None}& \cite{DeyPRB93-16} \\
\hline
    Ba$_3$IrTi$_2$O$_9$ & hexagonal & $P6_3mc$ & Ins. & -- & \multicolumn{2}{|c|}{None}&  \cite{DeyPRB86-12,CatuneanuPRB92-15} \\
    Ba$_3$ScIr$_2$O$_9$ & hexagonal & $P6_3/mmc$ & Ins.& -- & \multicolumn{2}{|c|}{None}& \cite{DeyPRB89-14} \\
    Ba$_3$YIr$_2$O$_9$ & hexagonal & $P6_3/mmc$ & Ins. & -- & FM & $T=4$\,K & \cite{DeyPRB89-14} \\
    Ba$_3$ZnIr$_2$O$_9$ & hexagonal & $P6_3/mmc$ & Ins. & -- & \multicolumn{2}{|c|}{None}& \cite{NagPRL116-16} \\
\hline
\hline
   \end{tabular}  
  \caption{ \label{tab-classIr} \small Main structural, transport and magnetic properties of Ir-based spin-orbit materials. In the third column, Ins. refers to {\it insulator} and MIT to {\it metal-insulator transition}. The notations AFM, FM and AIAO refer to a {\it antiferromagnetic}, {\it ferromagnetic} and {\it all-in-all-out} magnetic ordering respectively. }
  \end{center}
\end{table}

\begin{table}[!ht]
  \tiny
  \begin{center}
   \begin{tabular}{|c|c c|c l|c l|c|}
    \hline
    \hline
    Compound & \multicolumn{2}{|c|}{Crystal Struct.} & \multicolumn{2}{|c|}{Transport Property} & \multicolumn{2}{|c|}{Magnetic Ordering} & Ref. \\
\hline
\hline
    \multicolumn{8}{|c|}{\multirow{2}{*}{Ruthenium-based spin-orbit materials}} \\ 
    \multicolumn{8}{|c|}{} \\
\hline
\hline
     BaRuO$_3$ & cubic perovskite & $Pm\bar{3}m$ & Metal & & FM & $Tc=60$\,K & \cite{ZhouPRL101-08,JinPNAS-08,HanPRB93-16} \\
     CaRuO$_3$ & perovskite & $Pnma$ & Metal & & \multicolumn{2}{|c|}{None} & \cite{LongoJAP39-68,SchneiderPRL112-14,CapognaPRL88-02} \\
     SrRuO$_3$ & perovskite & $Pnma$ & Metal & & FM & $T_c=160$\,K & \cite{LongoJAP39-68,NoroJSPJ27-69,CapognaPRL88-02} \\
\hline
     Sr$_2$RuO$_4$ & K$_2$NiF$_4$-type & $I4/mmm$ & Metal & & \multicolumn{2}{|c|}{None} & \cite{FatuzzoPRB91-15,ZhangPRL116-16} \\
     Ca$_2$RuO$_4$ & distorted K$_2$NiF$_4$-type & $Pbca$ or $P2_1/c$ & MIT & $T=357$\,K & AFM & $T=110$\,K & \cite{AlexanderPRB60-99,MizokawaPRL87-01,LiuPRB88-13,FatuzzoPRB91-15} \\ 
\hline
     Sr$_2$RuO$_{6}$ & hexagonal & $P\bar{3}1m$ & Ins. & -- & AFM & $T_N=565$\,K & \cite{TianPRB92-15,HileyPRB92-15} \\
\hline
     Sr$_3$Ru$_2$O$_7$ & orthorhombic & $BBcb$ & Metal & & \multicolumn{2}{|c|}{None}  & \cite{CapognaPRL88-02,TamaiPRL101-08} \\
\hline
     Sr$_4$Ru$_3$O$_{10}$ & orthorhombic & $Pbam$ & Metal & & FM & $T_c =105$\,K & \cite{CapognaPRL88-02,MaoPRL96-06,GranataPRB93-16} \\
    \hline
    \hline
    \multicolumn{8}{|c|}{\multirow{2}{*}{Rhodium-based spin-orbit materials}} \\ 
    \multicolumn{8}{|c|}{} \\
\hline
\hline
     Li$_2$RhO$_3$ & honeycomb & $C 2/m$ & Ins. & gap: $0.08$\,eV & \multicolumn{2}{|c|}{None} & \cite{LuoPRB87-13,MazinPRB88-13} \\
\hline
     \srrho & distorted K$_2$NiF$_4$-type & $I4_1acd$ & Metal & & \multicolumn{2}{|c|}{None} & \cite{BaumbergerPRL96-06,PerryNJP9-06,MartinsPRL107-11} \\
\hline
     Sr$_4$RhO$_6$ & hexagonal & $R\bar{3}c$ & Ins. & gap: $0.1$\,eV & AFM & $T_N= - $\,K & \cite{CalderPRB92-15-2} \\
\hline
     Sr$_5$Rh$_4$O$_{12}$ &  & $P3c1$ & Ins. & -- & AFM & $T_N=23$\,K & \cite{CaoPRB75-07,CaoSSC141-07} \\
\hline
\hline
    \multicolumn{8}{|c|}{\multirow{2}{*}{Osmium-based spin-orbit materials}} \\ 
    \multicolumn{8}{|c|}{} \\
\hline
\hline
     BaOsO$_3$ & six-layer hexagonal 6H & $Pm\bar{3}m$ & Metal & & \multicolumn{2}{|c|}{None} & \cite{JungPRB90-14,ShiJACS135-13} \\
     CaOsO$_3$ & perovskite & $Pnma$ & Metal & & \multicolumn{2}{|c|}{None} & \cite{ShiJACS135-13} \\
     SrOsO$_3$ & perovskite & $Pnma$ & Metal & & \multicolumn{2}{|c|}{None} & \cite{ShiJACS135-13} \\
     NaOsO$_3$ & perovskite & $Pnma$ & MIT & $T=410$~K & AFM & $T=410$~K & \cite{PhysRevB.80.161104,JungPRB87-13,CalderPRL108-12,DuPRB85-12,
VecchioSciR3-13,CalderNatComm6-15} \\
\hline
     Cd$_2$Os$_2$O$_7$ & pyrochlore & $Fd\bar{3}m$ & MIT & $T=226$~K & AIAO & $T=226$~K & \cite{MandrusPRB63-01,PadillaPRB66-02,BogdanovPRL110-13,SohnPRL115-15} \\
\hline
     Ba$_2$NaOsO$_{6}$ & double-perovskite & $Fm\bar{3}m$ & Ins. & -- & FM & $T_c=6.8$\,K & \cite{LeeEPL80-07,SteelePRB84-11,GangopadhyayPRB91-15} \\ 
     Ba$_2$LiOsO$_{6}$ & double-perovskite & $Fm\bar{3}m$ & Ins. & -- & AFM & $T_N=8$\,K & \cite{SteelePRB84-11} \\
     Ba$_2$CaOsO$_{6}$ & double-perovskite & $Fm\bar{3}m$ & Ins. & --  & FM & $T_c=50$\,K & \cite{ThompsonJPCM26-14,GangopadhyayPRB93-16} \\
     Ba$_2$YOsO$_{6}$ & double-perovskite & $Fm\bar{3}m$ & Ins. & -- & AFM & $T_N=69$\,K & \cite{KermarrecPRB91-15,GangopadhyayPRB93-16} \\
    \hline
    \hline
   \end{tabular}  
  \caption{ \label{tab-classOs} \small Main structural, transport and magnetic properties of Ru,Rh and Os-based spin-orbit materials. In the third column, Ins. refers to {\it insulator} and MIT to {\it metal-insulator transition}. The notations AFM, FM and AIAO refer to a {\it antiferromagnetic}, {\it ferromagnetic} and {\it all-in-all-out} magnetic ordering respectively. }
  \end{center}
\end{table}

\subsection{Correlated spin-orbit insulators: the example of Sr$_2$IrO$_4$}

The 5d transition metal oxide (TMO) \sriro\ has a tetragonal crystal structure,
the symmetry of which is lowered from the K$_2$NiF$_4$-type, well-known 
in Sr$_2$RuO$_4$ or La$_2$CuO$_4$, by an $11^\circ$ rotation of its IrO$_6$ 
octahedra around the $\mathbf{c}$-axis \cite{HuangJSSC112-94}. Each Ir 
atom accomodates $5$ electrons and the standard picture neglecting
spin-orbit interactions would give a "t$_{2g}^5$" ground state. 
However, this compound exhibits insulating behavior up to the highest
measured temperatures, with a strongly temperature-dependent gap. 
The optical gap at room temperature is about $0.26$\,eV \cite{MoonPRB80-09}. 
Below $T_N=240$\,K, a canted-antiferromagnetic (AF) order sets in, 
with an effective 
local moment of $0.5$\,$\mu_B$/Ir, and a saturation moment of $0.14$\,$\mu_B$/Ir 
\cite{CaoPRB57-98}. This phase has triggered much experimental and 
theoretical work 
\cite{JinPRB80-09, WatanabePRL105-10,FujiyamaPRL112-14,LiuPRB92-15}, 
highlighting in particular the importance of the SOC.

Here, we focus on the paramagnetic phase, above $240$\,K, which is
most interesting due to the persistance of the insulating nature despite 
the absence of magnetic order, as shown by transport measurements 
\cite{KimScience323-09}, by scanning tunneling microscopy and spectroscopy 
experiments  \cite{DaiPRB90-14}, by angle-resolved spectroscopy 
\cite{KimPRL101-08, BrouetPRB92-15}, time-resolved
spectroscopy \cite{PhysRevB.93.241114, DHsieh-PRB86-12} 
or optical conductivity \cite{MoonPRB80-09}.

Resonant Inelastic X-ray spectroscopy (RIXS) experiments 
\cite{KimScience323-09} have early on proposed a picture in terms of
\jeff12\ and \j32\ states:

\begin{eqnarray}  \label{eq-jeff12}
 \left|j_\textrm{eff}=\frac{1}{2},m_{j_\textrm{eff}}=+\frac{1}{2}\right\rangle & = & +\displaystyle\frac{1}{\sqrt{3}}\Big(\left|d_{yz},\downarrow\right\rangle + i\left|d_{xz},\downarrow\right\rangle\Big) + \displaystyle\frac{1}{\sqrt{3}} \left|d_{xy},\uparrow \right\rangle \\
  \left|j_\textrm{eff}=\frac{1}{2},m_{j_\textrm{eff}}=-\frac{1}{2}\right\rangle & = & +\displaystyle\frac{1}{\sqrt{3}}\Big(\left|d_{yz},\uparrow\right\rangle - i\left|d_{xz},\uparrow\right\rangle\Big) - \displaystyle\frac{1}{\sqrt{3}} \left|d_{xy},\downarrow \right\rangle \nonumber
\end{eqnarray}

\begin{eqnarray} \label{eq-jeff32}
 \left|j_\textrm{eff}=\frac{3}{2},m_{j_\textrm{eff}}=+\frac{1}{2}\right\rangle & = & 
 \displaystyle-\frac{1}{\sqrt{6}}\Big(\left|d_{yz},\downarrow\right\rangle + i\left|d_{xz},\downarrow\right\rangle\Big) + \displaystyle\sqrt{\frac{2}{3}} \left|d_{xy},\uparrow\right\rangle \nonumber\\
 \left|j_\textrm{eff}=\frac{3}{2},m_{j_\textrm{eff}}=-\frac{1}{2}\right\rangle & = & 
 +\displaystyle\frac{1}{\sqrt{6}}\Big(\left|d_{yz},\uparrow\right\rangle - i\left|d_{xz},\uparrow\right\rangle\Big) + \displaystyle\sqrt{\frac{2}{3}} \left|d_{xy},\downarrow\right\rangle \\
 \left|j_\textrm{eff}=\frac{3}{2},m_{j_\textrm{eff}}=+\frac{3}{2}\right\rangle & = & \displaystyle -\frac{1}{\sqrt{2}}\Big(\left|d_{yz},\uparrow\right\rangle +i\left|d_{xz},\uparrow\right\rangle \Big) \nonumber\\
 \left|j_\textrm{eff}=\frac{3}{2},m_{j_\textrm{eff}}=-\frac{3}{2}\right\rangle & = & \displaystyle +\frac{1}{\sqrt{2}}\Big(\left|d_{yz},\downarrow\right\rangle -i\left|d_{xz},\downarrow\right\rangle \Big) \nonumber
\end{eqnarray}
Since the quartet of states lies lower in energy than the doublet and
the splitting between the \j32 and \jeff12 is large, neglecting any
band dispersion would result in a  
configuration with one electron in the \jeff12 
state. The DFT band structure displays a dispersion of width
comparable to this splitting, leaving the question {\it a priori} open again.
However, the bandwidth is narrowed due to structural distortions 
\cite{MartinsPRL107-11}, and electronic correlations can then become effective
and eventually drive the compound insulating. 

Since the discovery of this mechanism, other Ir-based compounds (cf. Tab.~\ref{tab-classIr}) have been classified as spin-orbit Mott insulators (Na$_2$IrO$_3$, pyrochlores, etc...). 
Recent theoretical studies predict also some fluoride material
\cite{BirolPRL114-15} to be in this class.
The one-orbital nature of insulating \sriro\ has contributed to 
intense activities attempting to dope the compound, with the hope
of inducing a superconducting state as in the cuprates.
Doping-induced metal-insulator transitions and the properties
of the metallic phases have therefore become a hot topic,
with studies of various compounds, e.g. 
\sriro \cite{BrouetPRB92-15,TorrePRL115-15}, 
(Sr$_{1-x}$La$_x$)$_3$Ir$_2$O$_7$ \cite{HoganPRL114-15}, 
Ca$_{1-x}$Sr$_x$IrO$_{3}$ \cite{ChengPRB83-11}, 
Ca$_{1-x}$Ru$_x$IrO$_{3}$ \cite{GunasekeraScRep5-15}, 
Sr$_2$Ir$_{1-x}$Rh$_x$O$_4$ \cite{QiPRB86-12,ChikaraPRB92-15}, 
Sr$_2$Ir$_{1-x}$Ru$_x$O$_4$ \cite{CalderPRB92-15}, 
Sr$_x$La$_{11-x}$Ir$_4$O$_{24}$ \cite{PhelanPRB91-15}.

\subsection{Correlated spin-orbit metals: the example of Sr$_2$RhO$_4$}

It is natural that also in {\it metallic} 4d or 5d transition metal compounds,
SOC can have notable consequences.
An example of a ``spin-orbit correlated metal'' is the end member SrIrO$_3$ 
of the Ir-based Ruddlesden-Popper Sr$_{n+1}$Ir$_n$O$_{3n+1}$ series 
\cite{MoonPRL101-08} but also many Ru-,Rh- or Os-based transition metal oxides (TMOs) belong to this
class (cf. Tab.~\ref{tab-classIr} and~\ref{tab-classOs}). 
In these compounds, correlations are important enough to 
renormalize the Fermi surface --
albeit in a strongly spin-orbit coupling-dependent way.
The respective roles of both effects
have been worked out in some details for several compounds, among which 
SrIrO$_3$ \cite{MoonPRL101-08,CaoPRB76-07,NiePRL114-15}, 
Sr$_2$RuO$_4$/Ca$_2$RuO$_4$ \cite{LiuPRB84-11,FatuzzoPRB91-15,ZhangPRL116-16} 
and \srrho \cite{LiuPRL101-08,HaverkortPRL101-08, MartinsPRL107-11}.

We will focus our attention in the following on \srrho\ motivated by
its structural proximity and isoelectronic nature to \sriro\ . 
Indeed, this TMO is the 4d counterpart of \sriro , both concerning
structure and filling. 
Understanding its Fermi surface requires to include both SOC and 
correlations \cite{MartinsPRL107-11}. It is composed of three pockets 
(cf. Fig.~\ref{fig7}): a circular hole-like $\alpha$-pocket around 
$\Gamma$, a lens-shaped electron pocket $\beta_M$ and a square-shaped 
electron pocket $\beta_X$ with a mass enhancement of $3.0$, $2.6$ and 
$2.2$ respectively \cite{BaumbergerPRL96-06}. 

In this review, 
we will put \sriro\ and \srrho\ in parallel, shedding light on the 
spectral properties of these compounds and elaborating on the 
notion of a reduced effective (spin-orbital) degeneracy that is
crucial for their properties.

\subsection{Spin-orbit coupling and cubic symmetry: the j$_\textrm{eff}$ picture}

Necessary conditions for realising a j$_\textrm{eff}$ picture are 
(1) a strong spin-orbit coupling constant and (2) an important cubic 
crystal field. 
These conditions are often met in
crystalline structures where IrO$_6$ octahedra
are present (cf. Tab.~\ref{tab-classIr}). 
Similar compounds based on Ru, Rh and Os also show such j$_\textrm{eff}$ states 
(cf. Tab~\ref{tab-classOs}). However, not all Ir-based structures belong to 
this case : we note that neither epitaxial thin films of 
IrO$_2$ \cite{KimPRB93-16} nor the correlated metal IrO$_2$ in its 
rutile structure \cite{PandaPRB89-14,KahkPRL112-14} exhibit such 
\jeff12\ state. 
We will now turn to a more precise description of that picture.

The spin-orbit interaction is one of the relativistic corrections to the 
Schr\"odinger-Pauli equation 
arising when taking the non-relativistic limit of Dirac's equation. 
It introduces a coupling between the spin $\mathbf{S}$ and the motion -- 
or more precisely the orbital momentum $\mathbf{L}$ in the atomic case -- 
of the electron. 
In a solid described within an independent-particle picture, spin-orbit 
coupling has the following general form:
\begin{equation} \label{eq-HSOgal}
 H_{SO}=\frac{\hbar}{4m_0^2c^2}{\boldsymbol \sigma}\cdot[{\boldsymbol \nabla} V(\mathbf{r})\times\mathbf{p}]
\end{equation}
where $m_0$ is the electron mass, $V(\mathbf{r})$ is the effective 
Kohn-Sham potential and $\sigma_{i=x,y,z}$ denote the Pauli-spin matrices.
Assuming that the potential close to the nucleus has spherical symmetry 
the mean value of the spin-orbit interaction on the atomic state $(n,\ell)$ takes the more common form:
\begin{equation} \label{eq-HSOatom}
 H_{SO}=\zeta_{SO}(n\ell)~\mathbf{l}\cdot\mathbf{s} \qquad \textrm{with} \qquad \zeta_{SO}(n\ell)= \frac{\hbar^2}{2m_0^2c^2}~\left\langle\frac{1}{r}\frac{dV}{dr}\right\rangle_{(n,\ell)}
\end{equation}
where $\mathbf{S}=\frac{1}{2}{\boldsymbol \sigma}$, $\mathbf{L}=\mathbf{r}\times\mathbf{p}=\hbar~\mathbf{l}$ and $\langle\dots\rangle_{(n,\ell)}$ denotes the mean value of the radial quantity 
in the state $(n,\ell)$. Tab.~\ref{tab-zetaSO} gives some values of the 
spin-orbit constant $\zeta_{SO}$ for $3d$, $4d$ and $5d$ atoms. The SOC increases with 
the atomic number, explaining why spin-orbit materials are mostly found 
in $5d$ and $4d$ TMOs.

\begin{table}[!ht]
  \small
  \begin{center}
   \begin{tabular}{|c|c|c||c|c|c||c|c|c|}
    \hline
    \hline
    Atom & Z & $\zeta_{SO}(3d)$ & Atom & Z & $\zeta_{SO}(4d)$ & Atom & Z & $\zeta_{SO}(5d)$\\
    \hline
    Fe & 26 & 0.050~eV & Ru & 44 & 0.161~eV & Os & 76 & 0.31~eV \\
    Co & 27 & 0.061~eV & Rh & 45 & 0.191~eV & Ir & 77 & 0.4~eV \\
    Cu & 29 & 0.103~eV & Ag & 47 & 0.227~eV & Au & 79 & 0.42~eV \\
    \hline
    \hline
   \end{tabular}  
  \caption{ \small Value of the spin-orbit constant $\zeta_{SO}$ in the 
$d$-valence shells of some transition metals. Data from Landolt-B\"ornstein database and \cite{Cotton2007} (3d), from \cite{HaverkortPRL101-08,Pitzer04} (4d) and from \cite{Friedel1969} (5d) \label{tab-zetaSO}}
  \end{center}
\end{table}

\begin{figure}
 \begin{center}
  \includegraphics[width=0.9\linewidth,keepaspectratio]{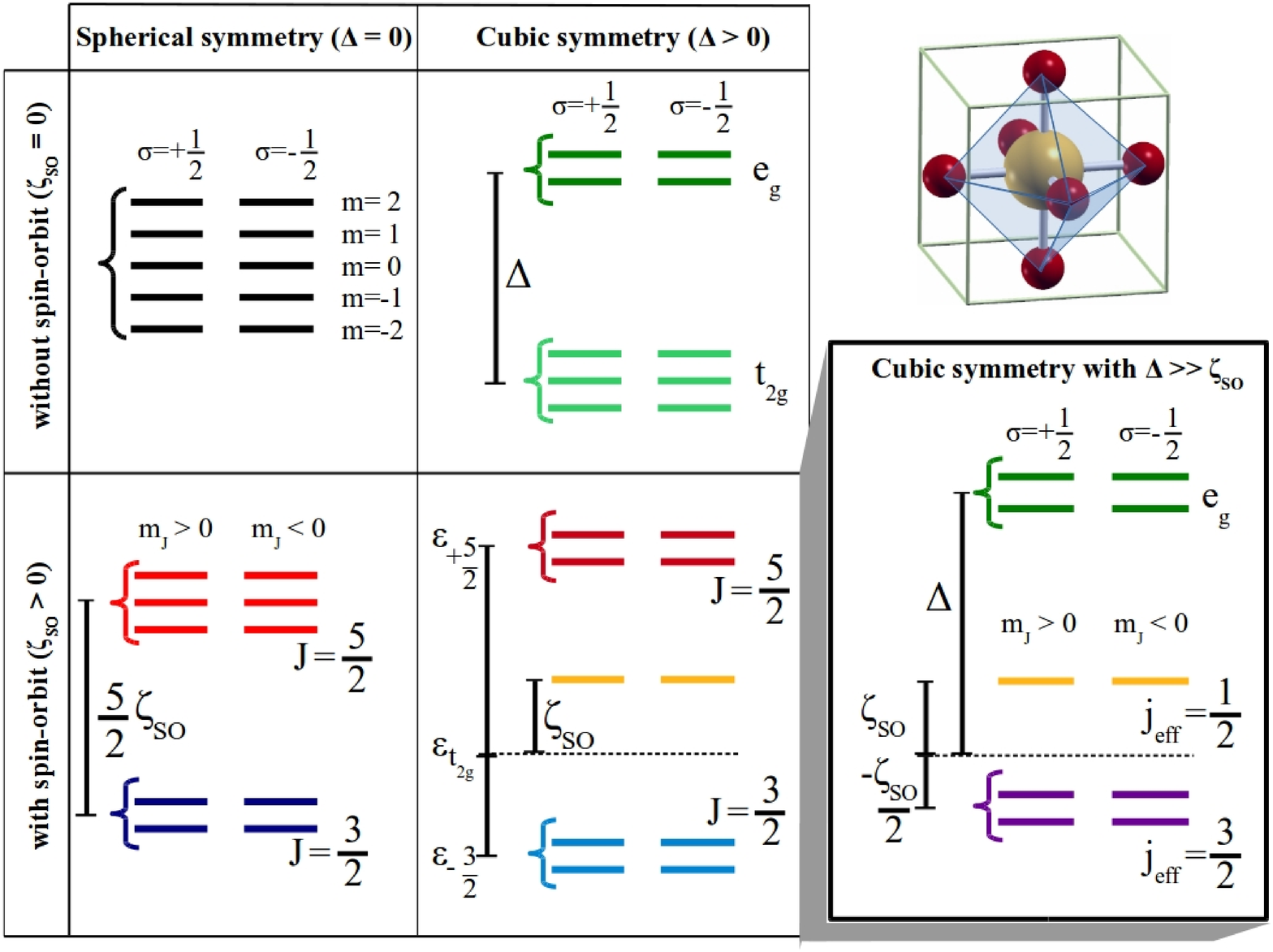}
 \end{center}
 \caption{\label{figtest} \small 
Orbital diagrams for the $d$-shell of an atom as a function of a cubic 
crystal field $\Delta$ and spin-orbit coupling $\zeta_{S0}$,
paramagnetic case. Starting from 
the $d$-shell in spherical symmetry, the cubic crystal field splits them 
into $e_g$ and $t_{2g}$, while the SOC creates a 6-fold $J=5/2$ multiplet and 
a $J=3/2$ quartet of lower energy. When both parameters are at stake, one 
gets a new multiplet structure where $J$ remains a good quantum number but
not $J_z$. 
The initial $J=5/2$ multiplet splits into a quartet and a doublet of lower 
energy, while the quartet $J=3/2$ undergoes some redefinition inside its 
submanifold. The energetic splitting and the nature of the spin-orbitals 
depend on the ratio between $\Delta/\zeta_{SO}$. An exception is the doublet 
which is already of the form of the \jeff12 . In the limit where 
$\Delta >> \zeta_{SO}$, as is the case in the compounds of our interest, 
one gets the celebrated splitting into $e_g$, \jeff12 and \j32 .} 
\end{figure}

Due to the effect of SOC, a multiplet splitting arises in the $d$-orbitals. 
Fig.~\ref{figtest} shows the multiplet splitting of $d$-orbitals due to 
the spin-orbit coupling as a function of the strength of a cubic crystal 
field $\Delta=10Dq$. 

In spherical symmetry the fine structure is composed of a six-fold 
$J=5/2$ multiplet (in red) and a $J=3/2$ quartet of lower energy (in blue), 
following "Land\'e's interval rule". The presence of a 
cubic crystal field splits further the six-fold multiplet. Indeed, the 
spin-orbit interaction in the cubic basis ($e_g$ and $t_{2g}$ in green and 
light green respectively in Fig.~\ref{figtest}) can be reduced to two 
five-dimensional submatrices:
\begin{equation}
 \left( \begin{array}{ccc|cc}
 0 & -i & i & \sqrt{3} & -1 \\
 i & 0 & -1 & -i\sqrt{3} & -i \\
 -i & -1 & 0 & 0 & -2i\\
 \hline
 \sqrt{3} & i\sqrt{3}& 0 & 0 & 0 \\
  -1 & i & 2i & 0 & 0 \\
 \end{array} \right).\frac{\zeta_{SO}}{2} \qquad \textrm{and} \qquad
 \left( \begin{array}{ccc|cc}
 0 & i & i & -\sqrt{3} & 1 \\
 -i & 0 & 1 & -i\sqrt{3} & -i\\
 -i & 1 & 0 & 0 & 2i\\ 
 \hline
 -\sqrt{3} & i\sqrt{3} & 0 & 0 & 0 \\
 1 & i & -2i & 0 & 0 \\
 \end{array} \right).\frac{\zeta_{SO}}{2} \\
\end{equation}
in the bases $\{d_{xz}\uparrow, d_{yz}\uparrow, d_{xy}\downarrow, d_{3z^2-r^2}\downarrow, d_{x^2-y^2}\downarrow\}$ and $ \{d_{xz}\downarrow, d_{yz}\downarrow, d_{xy}\uparrow, d_{3z^2-r^2}\uparrow, d_{x^2-y^2}\uparrow\}$ respectively. After diagonalization, the total angular momentum $J$ remains a good quantum number, contrary to $j_z/m_j$ and one gets the following fine structure:
\begin{itemize}
 \item a first quartet of $J=5/2$ states (in red) with an energy $$\varepsilon_{\frac{5}{2}+}=\frac{1}{4}\left(2\Delta-\zeta_{SO}\right) + \frac{1}{4}\sqrt{\left(2\Delta+\zeta_{SO}\right)^2+24 \zeta_{SO}^{~~~2}}$$
 \item a doublet of $J=5/2$ states (in yellow) of energy $$\varepsilon_{\frac{5}{2}-}= +2\frac{\zeta_{SO}}{2} $$
 \item a quartet of $J=3/2$ states (in light blue) with an energy $$\varepsilon_{\frac{3}{2}}=\frac{1}{4}\left(2\Delta-\zeta_{SO}\right) - \frac{1}{4}\sqrt{\left(2\Delta+\zeta_{SO}\right)^2+24 \zeta_{SO}^{~~~2}}$$
\end{itemize}
In the limit of strong crystal field ($\Delta>>\zeta_{SO}$), the $J=5/2$ 
doublet (in yellow) remains invariant while the higher-energy quartet will 
tend to the usual $e_g$ states and the lower-energy $J=3/2$ quartet will 
be composed of $t_{2g}$ states only, with an energy of $-\zeta_{SO}/2$. 

Since the SOC-matrix restricted to the \t2g subspace is exactly the opposite 
of the SOC-matrix of the $p$-states of a free atom, one usually labels these 
latter states by a $j_\textrm{eff}$ quantum number in analogy with the 
$p_\frac{1}{2}$ and $p_\frac{3}{2}$ multiplets, leading to the expressions 
given in Eq.~\eqref{eq-jeff12} and \eqref{eq-jeff32}. We point out that 
the \jeff12 doublet arises from the interplay of both cubic symmetry and 
SOC, whatever the strength of the crystal field. The corresponding 
eigenstates can indeed be written: 
\begin{eqnarray} 
 \left|\frac{1}{2},+\frac{1}{2}\right\rangle & = & +\displaystyle\frac{1}{\sqrt{3}}\Big(\left|d_{yz},\downarrow\right\rangle + i\left|d_{xz},\downarrow\right\rangle\Big) + \displaystyle\frac{1}{\sqrt{3}} \left|d_{xy},\uparrow \right\rangle = \frac{i}{\sqrt{6}}\Big(\sqrt{5}\left|\frac{5}{2},-\frac{3}{2}\right\rangle -\left|\frac{5}{2},\frac{5}{2}\right\rangle\Big) \nonumber\\
  \left|\frac{1}{2},-\frac{1}{2}\right\rangle & = & +\displaystyle\frac{1}{\sqrt{3}}\Big(\left|d_{yz},\uparrow\right\rangle - i\left|d_{xz},\uparrow\right\rangle\Big) - \displaystyle\frac{1}{\sqrt{3}} \left|d_{xy},\downarrow \right\rangle = \frac{i}{\sqrt{6}}\Big(\sqrt{5}\left|\frac{5}{2},\frac{3}{2}\right\rangle -\left|\frac{5}{2},-\frac{5}{2}\right\rangle\Big) \nonumber
\end{eqnarray}
(where the right-hand side is written using the $J,m_J$ quantum numbers). 
This may explain the robustness of this doublet in spin-orbit compounds 
\cite{CalderPRB86-12}. 
However, the splitting between the \jeff12 and \j32\ multiplets follows 
the inverse Land\'e interval rule (with the \jeff12\ above the \j32\ 
states) only in the strong crystal field limit.

\section{Interplay of spin-orbit interaction and Coulomb correlations from first principles}

\subsection{DFT+DMFT calculations with spin-orbit coupling \label{part-ldadmft}}

Combined density functional theory (DFT) and dynamical mean-field theory (DMFT), as
pioneered in \cite{LichtensteinPRB57-98, AnisimovJPCM9-97}
(for a review, see \cite{biermann_ldadmft,SilkeReview}), has made
correlated electron systems accessible to first principles calculations.
Over the years, various classes of systems ranging from 
transition metals~\cite{Mn,Fe,Braun,PhysRevB.82.104414}, 
their oxides~\cite{biermann:026404,tomczak_vo2_proc,optic_epl,optic_prb,
PoteryaevPRB76-07,tomczak_v2o3_proc,Thunstrom}, 
sulphides~\cite{lechermann:085101, lec05}, pnictides~\cite{AichhornPRB80-09,
MiyakeJPSJ79-10,AichhornPRB82-10,haule11},
rare earths~\cite{amadon:066402, cerium_zoelfl, PhysRevB.89.195132} and
their compounds~\cite{PhysRevB.76.235101, cefeaso, cefeaso2, TomczakPNAS110-13},
including heavy fermions~\cite{CeIrIn5, CeCu2Si2},
actinides~\cite{deltaPu, PhysRevLett.96.036404} 
and their compounds~\cite{PhysRevB.92.085125, PhysRevB.72.115106}
to organics \cite{organics}, correlated semiconductors \cite{PhysRevB.82.085104,PhysRevB.88.245203}, 
and correlated surfaces and interfaces~\cite{lao-sto,
PhysRevLett.110.166401, Hansmann_uncertainty}
have been studied with great success.
Besides intensive methodological developments (see e.g.~\cite{%
BiermannPRL90-03, ayral_gwdmft, ayralPRB87-13,
TomczakEPL100-12, TomczakPRB90-14, vanREPL108-14,
BiermannJPCM26-14}), recent research activities continue to
extend to new materials classes. In this context, also
4d and 5d oxides have come into focus
\cite{MartinsPRL107-11,AritaPRL108-12,AhnKunesJPCM27-15}. 
In this section, we review the technical aspects related to 
combined DFT+DMFT calculations in the presence of spin-orbit
interactions. Since the applications we later focus on are
4d and 5d oxides in their {\it paramagnetic} phases, we restrict
the discussion to this case. 

\begin{figure}
 \begin{center}
  \includegraphics[width=0.9\linewidth,keepaspectratio]{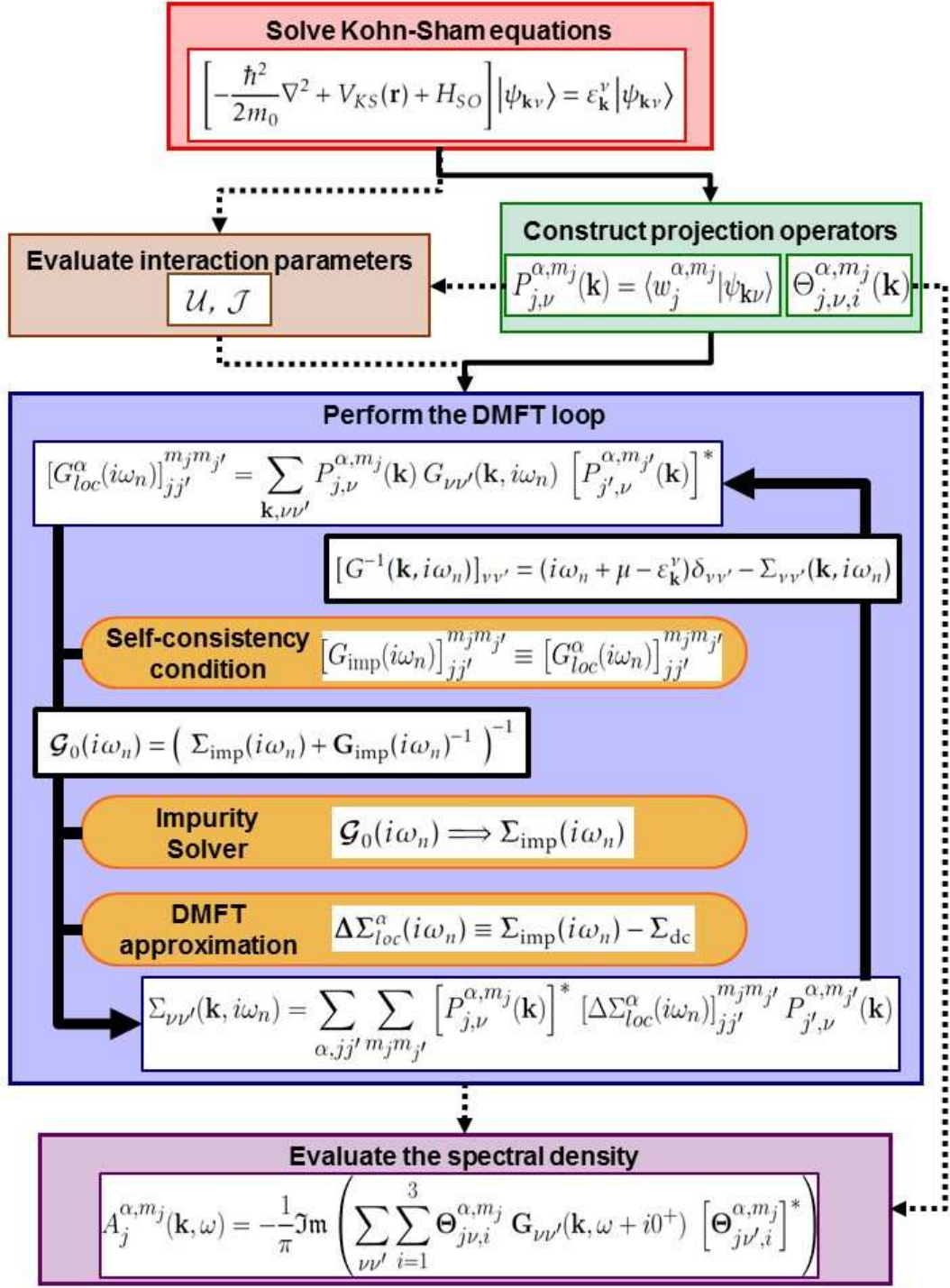}
 \end{center}
 \caption{\label{fig-DMFTcycle} \small Projector-based implementation of DFT+DMFT for calculations including spin-orbit coupling in the Kohn-Sham equations. Once the Kohn-Sham eigenstates $|\psi_{\mathbf{k}\nu}\rangle $ are known, their projections $P_{j,\nu}^{\alpha,m_j}(\mathbf{k})$ to the correlated Wannier-like orbitals $|w^{\alpha,m_j}_{j}\rangle$ are calculated. One can then build an effective local many-body atomic problem, subject to a self-consistency condition, which is solved using an impurity solver: this defines the DMFT loop (see Section~\ref{part-ldadmft}). The interaction parameters can also be evaluated consistenly using the projectors $P_{j,\nu}^{\alpha,m_j}(\mathbf{k})$ (see \cite{VaugierPRB86-12} and Section~\ref{part-Ueff}). After convergence of the DMFT cycle, the chemical potential is updated and the spectral function can be evaluated using partial projectors $\Theta_{j\nu,i}^{\alpha,m_j}(\mathbf{k})$ (see Appendix~A).} 
\end{figure}

In DMFT, a local approximation is made to the many-body self-energy
which can then be calculated from an effective atom problem, 
subject to a self-consistency condition (see Fig.~\ref{fig-DMFTcycle}).

The notion of locality is understood in the sense of
many-body theory as a site-diagonal form, with respect to
atomic sites after representing the Hamiltonian in an atom-centered
Wannier-type basis $|w^{\alpha,\sigma}_{\ell m}\rangle$, where the index $\alpha$ labels the atom in the unit-cell, $(\ell,m)$ the angular momentum quantum numbers of the atomic orbital and $\sigma$ the spin degree of freedom.
Different choices are possible for the construction of the
atom-centered orbitals, and the work reviewed here is based on
the construction of projected atomic orbitals subject to a 
subsequent orthonormalisation procedure \cite{AichhornPRB80-09}.

The DMFT self-consistency cycle links the local effective atom 
problem to the
electronic structure of the solid, via the transformation matrix 
from the Kohn-Sham states $|\psi_{\mathbf{k}\nu}^\sigma\rangle $, labelled 
by their momentum $\mathbf{k}$ their band index $\nu$ and their 
spin $\sigma$, to the resulting Wannier-like local orbitals 
$|w^{\alpha,\sigma}_{\ell m}\rangle$ .
These key quantities are called \textit{projectors} and denoted 
$ P_{\ell m,\nu}^{\alpha,\sigma}(\mathbf{k})$.

The main advantage of projector-based implementations of DFT+DMFT 
(see e.g. \cite{AnisimovPRB71-05,LechermannPRB74-06,AichhornPRB80-09})
is that not only the DFT-based part of the calculations but also the
determination of the local Green's function, used within the DMFT
self-consistency condition, can be performed in any convenient basis 
set, and notably in the one used in the respective DFT code.
Since the transformation of the 
DFT Hamiltonian matrix in that basis into the Kohn-Sham eigenset 
$|\psi_{\mathbf{k}\nu}^\sigma\rangle $ is known, it is sufficient to further 
determine the projections of the Kohn-Sham 
eigenstates onto the local orbitals $|w^{\alpha,\sigma}_{\ell m}\rangle$ 
used in the DMFT impurity problem. This is precisely the role of the projectors.

In \cite{MartinsPRL107-11}, this construction was generalised to the
case when spin is not a good quantum number any more, and
implemented within the framework of the DFT+DMFT implementation of
Ref.~\cite{AichhornPRB80-09}. Nowadays, it is available within
the TRIQS/DFTTools package \cite{dfttools} that links the 
Wien2k code \cite{Wien2kRef} to DMFT.
We give here the main lines of this generalisation of the 
projector-based DFT+DMFT formalism.

When taking into account SOC, the Kohn-Sham eigenstates 
$|\psi_{\mathbf{k}\nu}\rangle $ 
are built out of \textit{both} spin-up and spin-down states -- in a 
similar fashion as the previously introduced 
\jeff12\ and \j32\ atomic 
states. Nevertheless, we can still write them in the 
following Bloch form:

\begin{eqnarray} \label{eq-psiplusmoins}
 \psi_{\mathbf{k}\nu}(\mathbf{r}) & = & \left[ u^\uparrow_{\mathbf{k}\nu}(\mathbf{r}) +  u^\downarrow_{\mathbf{k}\nu}(\mathbf{r}) \right]e^{i\mathbf{k}\cdot\mathbf{r}} \nonumber\\
 & = & \phi^\uparrow_{\mathbf{k}\nu}(\mathbf{r}) +  \phi^\downarrow_{\mathbf{k}\nu}(\mathbf{r}).
\end{eqnarray}
where the index $\nu$ now runs over \textit{both} spin and band indices.
The state $|\phi_{\mathbf{k}\nu}^\sigma\rangle$ denotes the projection of the Kohn-Sham state 
onto its spin-$\sigma$ contribution and is \textit{not} an eigenstate of the Hamiltonian.

Using this decomposition, we can define the new projectors: 
\begin{equation} \label{eq-projSO}
 P_{\ell m,\nu}^{\alpha,\sigma}(\mathbf{k})=
 \langle w_{\ell m}^{\alpha,\sigma}|\psi_{\mathbf{k}\nu}\rangle = \langle w_{\ell m}^{\alpha,\sigma}|\phi_{\mathbf{k}\nu}^\sigma\rangle
\end{equation} 
We define them in the standard complex basis, but 
allow for a basis transformation to quantum numbers 
$j,m_j$ (like \jeff12\ and \j32) afterwards by means of
a unitary matrix transformation in the correlated $\ell$-space:
\begin{equation} \label{eq-projjmj}
 P_{j,\nu}^{\alpha,m_j}(\mathbf{k})= \sum_{m,\sigma} \mathcal{S}_{j,\ell m}^{m_j,\sigma}
 \langle w_{\ell m}^{\alpha,\sigma}|\psi_{\mathbf{k}\nu}\rangle = \sum_{m,\sigma} \mathcal{S}_{j,\ell m}^{m_j,\sigma}P_{\ell m,\nu}^{\alpha,\sigma}(\mathbf{k})
\end{equation}

The main difference with the usual implementation where spin is 
a good quantum number is that there are now two projectors associated 
to each band index $\nu$: 
$P_{\ell m,\nu}^{\alpha,\sigma}(\mathbf{k})$ with $\sigma=\uparrow,\downarrow$.

Using the decomposition 
\eqref{eq-psiplusmoins} in the formulation of the 
self-consistency condition relating the lattice Green's function 
of the solid to the impurity
model, the (inverse) Green's function of the solid is given by:
\begin{equation} \label{eq-GlattSO}
 [G^{-1}(\mathbf{k},i\omega_{n})]_{\nu\nu'} = (i\omega_{n} +\mu -\varepsilon_{\mathbf{k}}^{\nu}) \delta_{\nu\nu'} -\Sigma_{\nu\nu'}(\mathbf{k},i\omega_{n}),
\end{equation}
where $\varepsilon_{\mathbf{k}}^{\nu}$ are the ($\nu$-dependent only) 
Kohn-Sham eigenvalues and 
$\Sigma_{\nu\nu'}(\mathbf{k},i\omega_{n})$ is the approximation to 
the self-energy obtained by the solution of the DMFT impurity 
problem. It is obtained
by "mapping" the impurity self-energy to the local self-energy 
of the lattice and "upfolding" it as: 
\begin{equation} \label{eq-selfLDADMFT3}
\Sigma_{\nu\nu'}(\mathbf{k},i\omega_{n})=\sum_{\alpha,jj'}\sum_{m_jm_{j'}} \left[P_{j,\nu}^{\alpha,m_j}(\mathbf{k})\right]^*~\left[\Delta\Sigma^{\alpha}_{loc}(i\omega_{n})\right]^{m_jm_{j'}}_{jj'}~P_{j',\nu'}^{\alpha,m_{j'}}(\mathbf{k}).
\end{equation}
with 
\begin{equation}
 \left[\Delta\Sigma^{\alpha}_{loc}(i\omega_{n})\right]^{m_jm_{j'}}_{jj'} = \left[\Sigma_\mathrm{imp}(i\omega_n)\right]^{m_jm_{j'}}_{jj'}-\left[\Sigma_{dc}\right]_{jj'}^{m_jm_{j'}}
\end{equation}
Here, $\Sigma_\mathrm{imp}(i\omega_n)$ is the impurity self-energy, expressed in
the local orbitals, and $\Sigma_{dc}$ is the 
double-counting correction. 
Consequently, the equations of the DMFT loop (see Figure 3) 
are formally the same 
as in the case without SOC but 
the computations now involve matrices which are double in size.

The local Green's function is obtained by projecting the
lattice Green's function to the set of correlated orbitals 
and summing over the full Brillouin zone,
\begin{equation} \label{eq-Glatt3}
\left[G_{loc}^\alpha(i\omega_{n})\right]_{jj'}^{m_jm_{j'}}=\sum_{\mathbf{k},\nu\nu'} P_{j, \nu}^{\alpha,m_j}(\mathbf{k})~G_{\nu\nu'}(\mathbf{k},i\omega_{n})~\left[P_{j',\nu'}^{\alpha,m_{j'}}(\mathbf{k})\right]^*.
\end{equation}
In practice, the summation over momenta 
is done in the irreducible Brillouin zone only, 
supplemented by a standard symmetrization procedure, using 
Shubnikov magnetic point groups \cite{PhdMartins,Shubnikov51}.

The DMFT equations are solved iteratively:
starting from an initial local Green's function $G_{loc}^\alpha(i\omega_{n})$ (obtained from the "pure" Kohn-Sham lattice Green's function using Eq.~\eqref{eq-Glatt3}), the Green's function  $\mathcal{G}_0(i\omega_{n})$ of the effective environment in the impurity model is constructed. The impurity model is solved, allowing to evaluate the local self-energy of the solid (cf. Eq.~\ref{eq-selfLDADMFT3}) and a new lattice Green's function $G(\mathbf{k},i\omega_{n})$. The latter can then be projected again onto the correlated subset and the cycle is repeated until convergence is reached.

\subsection{Computation of the Wannier projectors within the augmented plane wave framework}

The present implementation is within a full-potential linearized
augmented plane wave (FLAPW) framework, as realised in the Wien2k
package \cite{Wien2kRef}. With respect to the existing DFT+DMFT implementation
\cite{AichhornPRB80-09} in this context, 
the main changes concern the projection technique for building 
the correlated orbitals: as discussed above,
one has to take care of the fact that
spin is no longer a good quantum number, leading to the more general
construction of localized ``spin-orbitals''.
The necessary modifications in the construction of the projectors 
are reviewed in the following.

As in the case without SOC, we still use the Kohn-Sham 
states within a chosen energy window
$\mathcal{W}$ to form the Wannier-like functions that are treated as 
correlated orbitals, and the construction of the Wannier projectors
is done in two steps.
First, auxiliary Wannier projectors 
$\widetilde{P}_{\ell m,\nu}^{\alpha,\sigma}(\mathbf{k})$ 
are calculated -- separately for each $|\phi^\sigma_{\nu\mathbf{k}}\rangle$ 
term -- from the following expression:
\begin{eqnarray} \label{eq-PtildeAlmSO}
\widetilde{P}_{\ell m,\nu}^{\alpha,\sigma}(\mathbf{k}) & = & \langle u_\ell^{\alpha,\sigma}(E_1\ell)Y^\ell_m |\psi_{\mathbf{k}\nu}\rangle \nonumber\\
& = & A^{\nu\alpha}_{\ell m}(\mathbf{k},\sigma) +\overset{N_{LO}}{\underset{n_{LO}=1}{\sum}}
c_{LO}^{\nu,\sigma} C_{\ell m}^{\alpha,LO}\mathcal{O}_{\ell m,\ell'm'}^{\alpha,\sigma}.
\end{eqnarray}
A description of the augmented plane wave (APW) basis can be found in 
Ref. \cite{AichhornPRB80-09}. We use the same notations 
e.g. for the coefficients $A^{\nu\alpha}_{\ell m}(\mathbf{k},\sigma)$ 
and the overlap matrix $\mathcal{O}_{\ell m,\ell'm'}^{\alpha,\sigma}$ 
as introduced there.

One performs an orthonormalisation step
in order to get the Wannier projectors 
$P_{\ell m,\nu}^{\alpha,\sigma}(\mathbf{k})$. 
The overlap matrix 
$\left[O(\mathbf{k})\right]^{\alpha,\alpha'}_{(m\sigma),(m'\sigma')}$ 
between the correlated $\ell$ orbitals is defined by:
\begin{eqnarray}
\left[O(\mathbf{k})\right]^{\alpha,\alpha'}_{(m\sigma),(m'\sigma')} & = & 
\sum_{\nu=\nu_{min}(\mathbf{k})}^{\nu_{max}(\mathbf{k})}
\widetilde{P}_{\ell m,\nu}^{\alpha,\sigma}(\mathbf{k}) \widetilde{P}_{\ell m',\nu}^{\alpha',\sigma' *}(\mathbf{k}).
\end{eqnarray}
leading to the final projectors:
\begin{equation}
  P_{\ell m,\nu}^{\alpha,\sigma}
  (\mathbf{k})=\underset{\alpha',m',\sigma'}{\sum}\left
    \{ \left[O(\mathbf{k})
    \right]^{-1/2}\right\}^{\alpha,\alpha'}_{(m\sigma),(m'\sigma')}
    \widetilde{P}_{\ell m'\nu}^{\alpha',\sigma'}(\mathbf{k}),
\label{eq:wannier-proj}
\end{equation}
which are then further transformed into a $j,m_j$ basis as described above (cf. Eq.~\eqref{eq-projjmj}).

\subsection{Effective local Coulomb interactions from first principles \label{part-Ueff}}

Hubbard interactions $U$ -- obtained as the static ($\omega=0$) limit 
of the onsite matrix element
$\langle |W^{\rm partial} | \rangle$ within the
``constrained random phase approximation'' (cRPA) 
-- have by now been
obtained for a variety of systems, ranging from transition
metals \cite{AryasetiawanPRB70-04} to oxides
\cite{MiyakePRB77-08,AryasetiawanPRB74-06,TomczakPRB81-10,VaugierPRB86-12,
PhysRevB.87.165118},
pnictides \cite{MiyakeJPSJ77-08,NakamuraJPSJ77-08,MiyakeJPSJ79-10,ImadaJPSJ79-08}, 
f-electron elements \cite{PhysRevB.88.125123}
and compounds \cite{TomczakPNAS110-13}, 
to surface systems \cite{0953-8984-25-9-094005},
and several implementations
within different electronic structure codes and basis sets
have been done, e.g. within 
linearized muffin tin orbitals \cite{AryasetiawanPRB70-04, PhysRevB.71.045103}, maximally
localized Wannier functions \cite{MiyakePRB77-08,FriedrichPRB83-11,NakamuraJPSJ77-08} (as elaborated in \cite{mar97}), or
localised orbitals constructed from projected atomic
orbitals \cite{VaugierPRB86-12}.
The implementation into the framework of the Wien2k package
\cite{VaugierPRB86-12} made it possible that Hubbard $U$'s be calculated
for the same orbitals as the ones used in subsequent DFT+DMFT 
calculations, and, to our knowledge, Ref.~\cite{MartinsPRL107-11} was indeed
the first work using in this way consistently calculated Hubbard 
interactions in a DFT+DMFT calculation.
Systematic calculations investigating the basis set dependence
for a series of correlated transition metal oxides revealed
furthermore interesting trends, depending on the choice of the
low-energy subspace. In contrast to common belief until then,
Hubbard interactions increase for example with the principal quantum 
number when low-energy effective models encompassing only the t$_{2g}$
orbitals are employed. These trends can be rationalised by two
counteracting mechanisms, the increasing extension of the orbitals
with increasing principal quantum number and the less efficient
screening by oxygen states \cite{VaugierPRB86-12}.
We will come back to this point below, in the context of the
cRPA calculations for our target compounds.
\\

In the following, we review the specificities involved when determining
the Hubbard interactions for our target spin-orbit compounds. We hereby
use the same notations as in \cite{VaugierPRB86-12}.

We start from the standard Hubbard-Kanamori Hamiltonian $H_{int}$ which 
allows us to describe the interactions between \t2g\ orbitals within 
a Hamiltonian restricted to the
\t2g\ -space:
\begin{eqnarray} \label{eq-Hints}
  H_\textrm{int} &=& \mathcal{U} \sum_m n_{m\uparrow}  n_{m\downarrow} +
\mathcal{U}'\sum_{m<n,\sigma} n_{m\sigma} n_{n\bar{\sigma}} \nonumber \\
      &+& (\mathcal{U}'-\mathcal{J}) \sum_{m<n,\sigma} n_{m\sigma}n_{n \sigma} \\
    &-& \mathcal{J} \sum_{m<n,\sigma}\left[ c_{m\sigma}^{\dagger}
    c_{m\bar{\sigma}} c_{n\bar{\sigma}}^{\dagger} c_{n\sigma} + c_{m\sigma}^{\dagger}
    c^{\dagger}_{m\bar{\sigma}} c_{n\sigma} c_{n\bar{\sigma}}\right] \nonumber
\label{eq:hami}
\end{eqnarray}
where $\mathcal{U}$ is the intra-orbital Coulomb repulsion term and $\mathcal{U}'$ ($=\mathcal{U}-2\mathcal{J}$ with cubic symmetry) the interorbital Coulomb interaction which is reduced by Hund's exchange $\mathcal{J}$. ($m$ and $n$ run over the three \t2g\ orbitals and 
$\sigma$ stands for the spin). 

To draw the link between the cRPA calculations and this model Hamiltonian, 
the terms $\mathcal{U}$, $\mathcal{U}'$ and $\mathcal{J}$ are understood 
as the Slater-symmetrized effective interactions in the \t2g\ subspace, 
related to the Slater integrals $F^0$, $F^2$ and $F^4$ as:
\begin{eqnarray}
 \mathcal{U}=F^0+\frac{4}{49}(F^2+F^4) & \textrm{and} & \mathcal{J}=\frac{3}{49}F^2+\frac{20}{441}F^4 \quad 
\end{eqnarray}
The last relation $\mathcal{U}'=F^0-\frac{2}{49}F^2-\frac{4}{441} F^4$ is 
redundant since $\mathcal{U}'=\mathcal{U}-2\mathcal{J}$. 

One now transforms $H_\textrm{int}$ into the $j_{\textrm{eff}}$ basis using the unitary matrix transformation $\mathcal{S}_{j,lm}^{m_j,\sigma}$. 
Keeping only density-density terms, $H_\textrm{int}$ becomes:
\begin{equation}
 H_\textrm{int}=\frac{1}{2}\sum_{j,m_j}\sum_{j',m_{j'}} U^{m_jm_{j'}}_{jj'} n_{j,m_j} n_{j,m_{j'}} 
\end{equation}
Here, the index $j$ is a shortcut notation for the $j_\textrm{eff}=\{3/2,1/2\}$ 
quantum number and $m_j=\{\pm3/2,\pm1/2\}$. 
The reduced interaction matrix $U^{m_jm_{j'}}_{jj'}$ has the following form: 
\begin{equation}\label{eq-Ueff}
 U^{m_j m_{j'}}_{jj'}=U^{\overline{m}_j \overline{m}_{j'}}_{jj'}=\left(\begin{array}{c|cc}
               0 & \mathcal{U}-2\mathcal{J} & \mathcal{U}-\frac{5}{3}\mathcal{J} \\ 
               \hline
              \mathcal{U}-2\mathcal{J} & 0 & \mathcal{U}-\frac{7}{3}\mathcal{J} \\
               \mathcal{U}-\frac{5}{3}\mathcal{J} & \mathcal{U}-\frac{7}{3}\mathcal{J} & 0 \\
         \end{array}\right)
\end{equation}

\begin{equation}
 U^{m_j\overline{m}_{j'}}_{jj'}=U^{\overline{m}_jm_{j'}}_{jj'}=\left(\begin{array}{c|cc}
               \mathcal{U}-\frac{4}{3}\mathcal{J} & \mathcal{U}-\frac{8}{3}\mathcal{J} & \mathcal{U}-\frac{8}{3}\mathcal{J} \\ 
               \hline
               \mathcal{U}-\frac{8}{3}\mathcal{J} & \mathcal{U}-\mathcal{J} & \mathcal{U}-\frac{7}{3}\mathcal{J} \\
               \mathcal{U}-\frac{8}{3}\mathcal{J} & \mathcal{U}-\frac{7}{3}\mathcal{J} & \mathcal{U}-\mathcal{J} \\
         \end{array}\right)
\end{equation}

We use the standard convention that $\overline{m}_j$ denotes $-m_j$, as 
usually done for spin degree of freedom. The ordering of the orbitals 
$|j,|m_j|\rangle$ is: $|1/2,1/2\rangle,|3/2,1/2\rangle,|3/2,3/2\rangle$, 
\jeff12\ and \j32\ blocks are emphasized to ease the reading of the matrices. 

\subsection{Technicalities of the DMFT calculation}

For the solution of the quantum impurity problem we
apply the continuous-time quantum Monte Carlo method (CTQMC)
in the strong-coupling formulation \cite{WernerPRL97-06}. 
We are able to perform calculations at room 
temperature ($\beta=1/k_BT=40$~eV$^{-1}$) 
with reasonable numerical effort.
In our calculations, we use typically around $16\times 10^6$ 
Monte Carlo sweeps and 28 \textbf{k}-points in the irreducible Brillouin zone. 

Since the CTQMC solver computes the Green's function on the imaginary-time
axis, an analytic continuation is needed in order to obtain
results on the real-frequency axis. A continuation of the 
impurity self-energy using a stochastic version of the maximum 
entropy method \cite{BeachArXiv-04} yields real and imaginary parts 
of the retarded self-energy.
From those, we calculate the momentum-resolved spectral function 
$A(\mathbf{k},\omega)$ using partial projectors introduced in Appendix A.

During the calculations we use the Fully Localized 
Limit (FLL) expression for the double-counting:
\begin{equation}
 \Sigma_{j,j'}^{dc}=\left[ U(N_c-\frac{1}{2})-J(\frac{1}{2}N_c-\frac{1}{2}) \right]\delta_{jj'}
\end{equation}
where $j$ and $j'$ run over the $j_{\textrm{eff}}$ states and $N_c$ is the total occupancy 
of the orbitals. (Since each orbital is doubly degenerate in $m_j$, 
$N_c/2$ is used in the term 
containing $J$). Moreover, we neglect the off-diagonal terms 
in the local Green's functions (particularly, we neglect the term between the 
\jeff12\ and the \j32\ $|m_j|=1/2$ which we checked to be two
orders of magnitude smaller than the diagonal terms,
in the chosen basis).

\section{Electronic structure of Sr$_2$IrO$_4$ and Sr$_2$RhO$_4$}

\subsection{Electronic structure of Sr$_2$IrO$_4$ and Sr$_2$RhO$_4$ within DFT-LDA}

The Kohn-Sham band structures of \sriro\ and \srrho\ within the local 
density approximation and in the presence of spin-orbit coupling 
(LDA+SO) are represented in Fig.~\ref{fig2}-(d) and (e). For \sriro , 
we use the lattice parameters measured at 295~K in \cite{CrawfordPRB49-94}, 
and  for \srrho\ those measured at 300~K in \cite{VogtJSSC123-96}. 

\begin{figure}
 \begin{center}
  \includegraphics[width=0.9\linewidth,keepaspectratio]{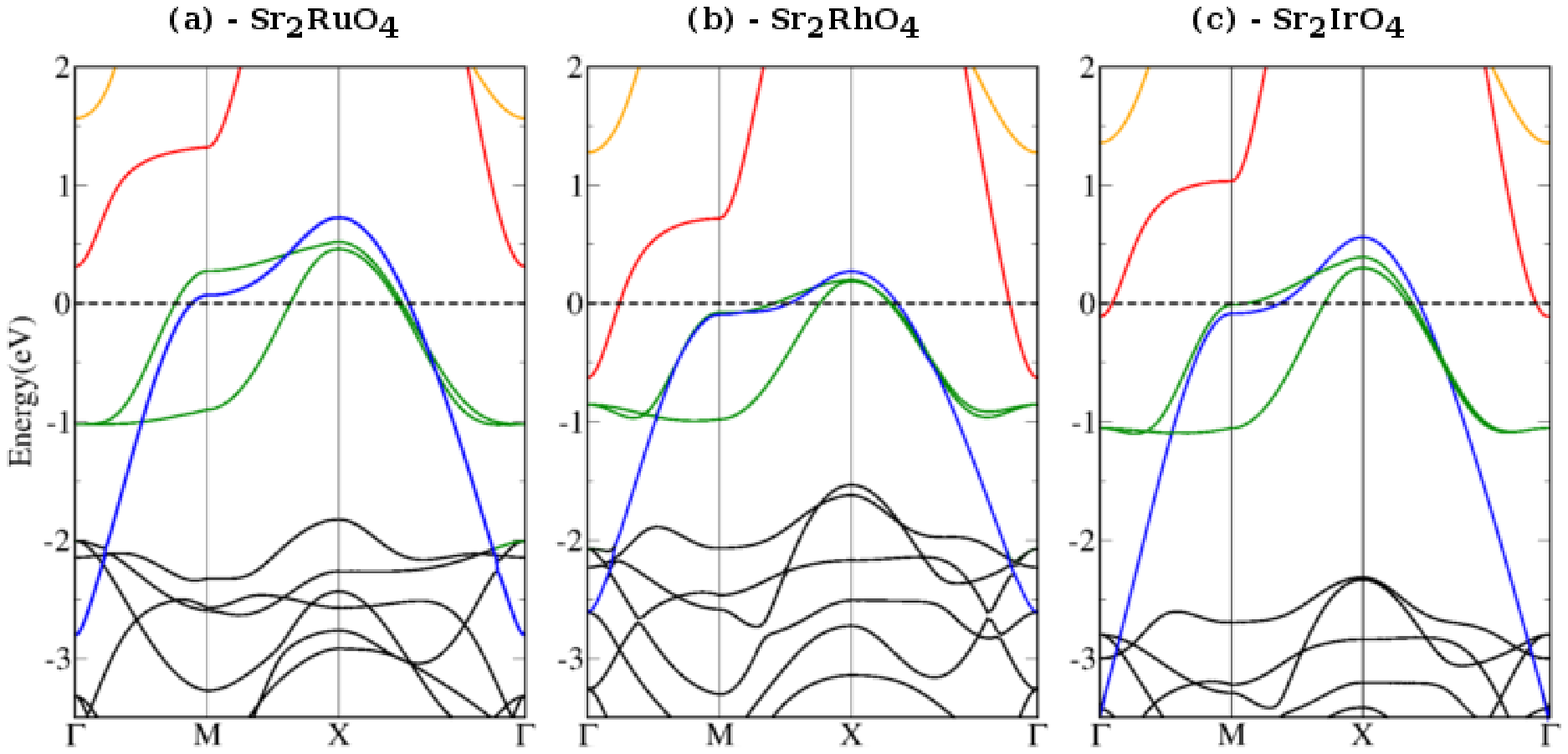}
 \end{center}
 \caption{\label{fig1} \small 
Kohn-Sham band structures of Sr$_2$RuO$_4$ (a) \srrho\ (b) and \sriro\ (c) within LDA (and without spin-orbit coupling), artificially assuming that both \srrho\ and \sriro\ crystallize in the same K$_2$NiF$_4$ structure as their Ru-counterpart. For Sr$_2$RuO$_4$, we use the lattice parameters at 300~K given in \cite{ChmaissemPRB57-98}. 
The t$_{2g}$-dominated bands are plotted in green ($d_{xy}$) and blue ($d_{xz}$ and $d_{yz}$) while the e$_g$ bands are in red ($d_{x^2-y^2}$) and yellow ($d_{3z^2-r^2}$), the O-$2p$ states in black. 
} 
\end{figure}

\begin{figure}
 \begin{center}
  \includegraphics[width=0.6\linewidth,keepaspectratio]{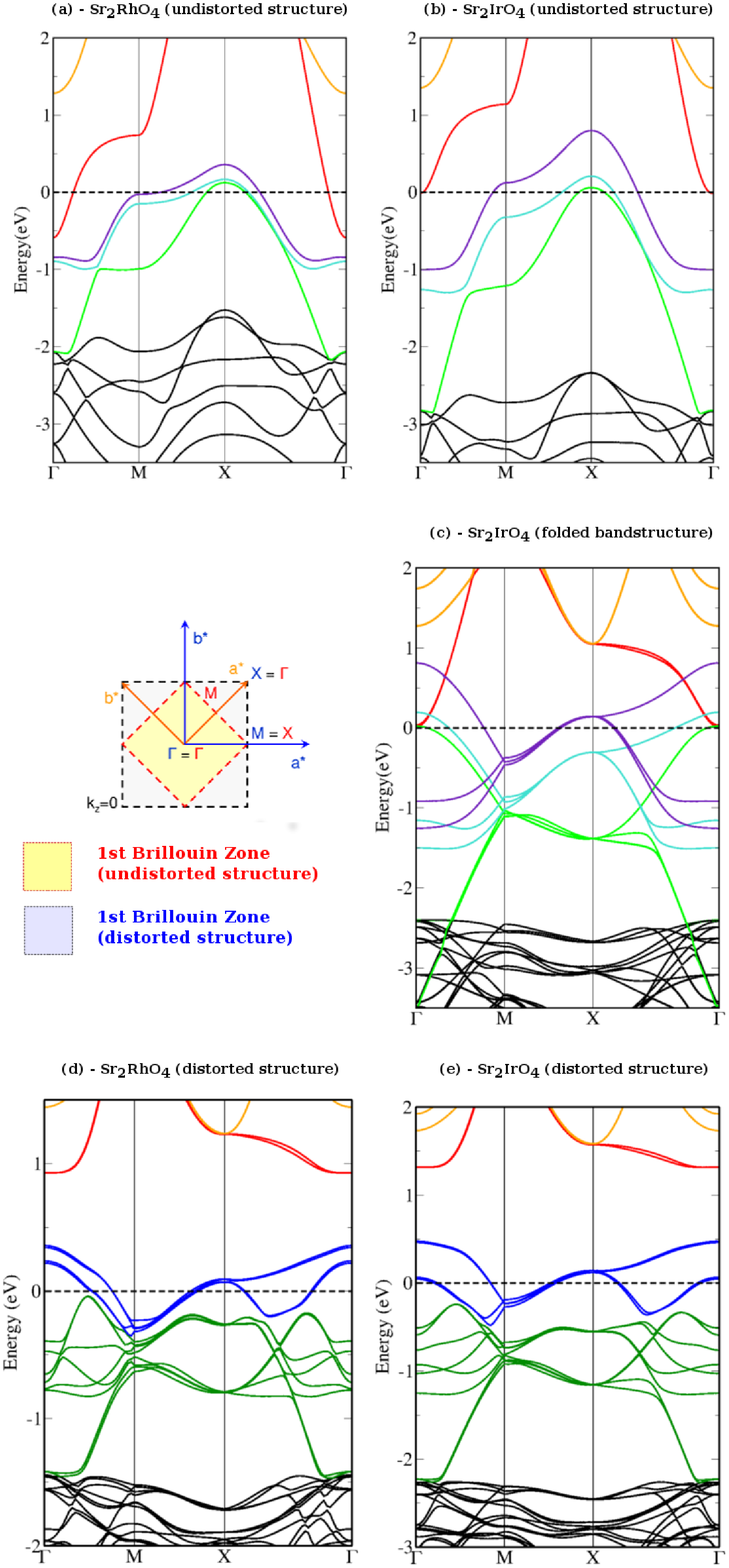}
 \end{center}
 \caption{\label{fig2} \small
Kohn-Sham band structures within LDA+SO of \srrho\ and \sriro\ assuming that they crystallize without distortions in a K$_2$NiF$_4$ structure (a-b), of \sriro\ in a supercell containing 4 "undistorted" unit-cells (c) and of "real" \srrho\ and \sriro\ (d-e).
The reduction of the first Brillouin zone, when the crystal symmetry is lowered, is also shown.
The e$_g$ states are in yellow ($d_{3z^2-r^2}$) and red ($d_{x^2-y^2}$).
In the t$_{2g}$ manifold, in purple the \jeff12 in light blue the \j32 $m_j=3/2$ and in green the \j32 $m_j=1/2$. In black, the O-$2p$ states.
} 
\end{figure}

The LDA+SO band structures for \sriro\ and \srrho\ are very similar, 
as a consequence of both, the structural similarity and the key role of  
spin-orbit coupling in these compounds. The e$_g$-states ($d_{x^2-y^2}$ in 
red and $d_{3z^2-r^2}$ in yellow) start at about 1 to 1.5 eV, and are fully 
separated from the t$_{2g}$-manifold which lies around the Fermi level 
and overlaps at lower energies with the oxygen $2p$-states (black). 
Given the t$_{2g}^5$ filling and the four-atom unit cell of both compounds, 
a  metallic solution is obtained within LDA for both \srrho\ and 
\sriro\ -- at variance with experiments for \sriro . 
Among the t$_{2g}$-manifold (in green), only the four highest-lying 
bands, highlighted in blue, cross the Fermi level: this is suggestive of 
the existence of a separated half-filled \jeff12 -derived band, which -- 
within a four-atom unit cell -- corresponds to a quartet of bands at 
each k-point. We stress however that the true picture is much more 
subtle: in fact, \jeff12 and \j32 overlap (see the band structure
between the $\Gamma$ and the M-point for instance) and the identification of the upper four bands as 
the \jeff12 states is too simplistic. 
We will come back to this point below.

To get a better understanding of the Kohn-Sham band structures of 
\srrho\ and \sriro , we 
study artificial compounds
where both the structural distortions and the spin-orbit coupling 
have been switched off.
Fig.~\ref{fig1}-(b) and (c) depict the LDA band structure of such 
"idealized undistorted \srrho\ and \sriro ". Neglecting the rotation 
of about $10^\circ$ of their IrO$_6$ and RhO$_6$ octahedra around the 
$c$-axis leads to a K$_2$NiF$_4$-type crystal structure, like in
Sr$_2$RuO$_4$, the well-known LDA band structure of which is plotted in  
Fig.~\ref{fig1}-(a).  

The similarity of the three band structures is obvious. 
Around the Fermi level, one distinguishes the three t$_{2g}$ bands. 
The $d_{xy}$-band (green) reaches out to lower energies and 
overlaps with the oxygen $2p$-states (black). The e$_g$-states 
($d_{x^2-y^2}$ in red and $d_{3z^2-r^2}$ yellow), higher in energy, cut the 
Fermi level in both \sriro\ and \srrho\ due the additional electron 
remaining in the $d$-manifold, contrary to \srruo , which has actually 
a mere t$_{2g}^4$-filling. The larger extension of the $5d$ orbitals (cf. 
Fig~\ref{fig0}) explains the wider bandwidth observed for \sriro\ in 
comparison to \srrho : the $d_{xy}$ band reaches the value of 
$-3.5$~eV in $\Gamma$, while it remains above $-3$~eV for the 4d 
counterparts. Another consequence of this wider extension is the 
stronger hybridization between the $5d$ states with the oxygen 
$p$-states, which are located 1~eV lower in energy in \sriro\ 
than in the $4d$-TMOs.

Re-introducing the effects of 
the spin-orbit coupling in \srrho\ and \sriro\ (but without considering
the structural distortions) modifies these Kohn-Sham 
band structures to those shown in Fig~\ref{fig2}-(a) and (b). 
The t$_{2g}$ bands are the most affected, while the e$_g$ bands are 
slightly shifted as a consequence of the topological change in the 
t$_{2g}$ manifold. A detailed study of the character of these band 
structures confirms the decoupling between e$_g$ and t$_{2g}$ states
(see also Refs. \cite{LiuPRL101-08, HaverkortPRL101-08}). 
The cubic crystal field at stake in these compounds is indeed much
larger than the energy scale associated to the spin-orbit coupling 
of about $\zeta_{\textrm{SO}}\approx 0.4$~eV and 
$\zeta_{\textrm{SO}}\approx 0.2$~eV for \sriro\ and \srrho\ respectively. 

The $j_\textrm{eff}$ picture is thus justified in both, \sriro\ and \srrho : 
the t$_{2g}$ orbitals split into a quartet of \j32 states and a higher 
lying doublet \jeff12 . Each state is doubly degenerate in $\pm m_j$ -- since 
we observe the system in its paramagnetic phase at room temperature and 
the crystal structure has a center of inversion. Therefore we  
still refer to them as the "\jeff12 band" and the two "\j32 bands" in 
the following.
The three $j_{\textrm{eff}}$ bands can easily be identified: the \jeff12\ 
one (light green) 
lies above the two \j32\ ones ($m_j=3/2$ in light blue and $m_j=1/2$ in 
violet). 
The three $j_\textrm{eff}$ bands are well-separated all along the 
$\mathbf{k}$-path, and more generally in the whole Brillouin zone. 
Since the spin-orbit coupling is twice smaller in \srrho , the splitting 
between the $j_\textrm{eff}$ bands is reduced by a factor of 2, as one 
can see for instance at $X$ or $\Gamma$.

To draw the link between the "undistorted" band structures and 
the realistic ones, we plot in Fig~\ref{fig2}-(c) the LDA+SO band 
structure of the \textit{undistorted} \sriro\ in a supercell 
containing four unit cells. Each band is now folded four times and 
we provide a scheme of the two first Brillouin Zones in the 
$\mathbf{k}_z=0$ plane to understand the correspondence between 
the high-symmetry points of each structure.  

Comparing Fig~\ref{fig2}-(c) and (e) highlights the key role of 
the structural distortion in \sriro :
an hybridisation between two neighboring Ir $d_{xy}$ and $d_{x^2-y^2}$ 
orbitals via the in-plane oxygens is now allowed and pushes the 
t$_{2g}$ and e$_g$ bands apart. Another consequence of the distortions 
is the general narrowing of the $j_\textrm{eff}$ bandwidth, which 
is of crucial importance to drive the compound 
insulating, as we will see below. 

Finally, comparing Fig~\ref{fig2}-(c) and (e) gives more insights 
into the nature of the four highest-lying bands (blue) of 
Fig~\ref{fig2}-(e). While along the $M-X$ direction, each quartet 
of $j_\textrm{eff}$ bands remain well-separated, \jeff12 and \j32 
overlap in the other direction $\Gamma-M$ and $M-X$. As a result, 
the \j32 bands cross the Fermi-level closest to the $\Gamma$-point, 
while the other crossings are due to the \jeff12 bands. The 
identification of the upper four bands in \sriro\ as "pure" 
\jeff12 states is thus too simplistic, implying the need for 
a Hamiltonian containing more than one orbital in a realistic
calculation.

The same mechanisms are at stake in \srrho\ even though we do not
display the orbital characters here:
the four highest-lying bands, highlighted in blue in Fig.~\ref{fig2}-(d) 
exhibit a mixed character of type \jeff12 and \j32 . Moreover, thanks to 
the distortions which allow the opening of a gap between t$_{2g}$ and e$_g$ 
bands, the LDA+SO Fermi surface becomes qualitatively similar to the 
experimental one: as shown in Fig.~\ref{fig7}-(a), they both contain 
three closed contours : a circular hole-like $\alpha$-pocket around 
$\Gamma$, a lens-shaped electron pocket $\beta_M$ and a square-shaped 
electron pockets $\beta_X$. However the striking discrepancies in the 
size of the pockets point out a subtle deficiency of the LDA for 
\srrho \cite{LiuPRL101-08, HaverkortPRL101-08}.

\subsection{Wannier functions}

We have derived the Wannier functions associated to the  
$j_{\textrm{eff}}$ manifold for both \sriro\ and \srrho , 
using the framework introduced in section \ref{part-ldadmft}. 
Because of the mixed character of the four bands that cross the Fermi level in \sriro\ and \srrho , the local effective atomic problem used in the DMFT cycle must contain the three $j_\textrm{eff}$ orbitals and thus accomodate five electrons. We construct Wannier functions for the $j_\textrm{eff}$ orbitals from the LDA+SO band structure of \sriro\ and \srrho , 
using an energy window $[-3.0,0.5]$\,eV for \sriro\ and an energy window $[-2.67;0.37]$\,eV for \srrho . 

\begin{figure}
 \begin{center}
  \includegraphics[width=\linewidth,keepaspectratio]{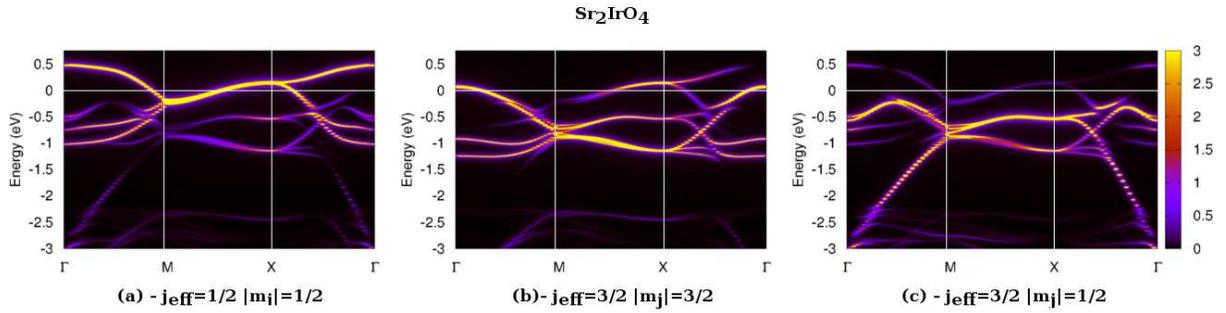}
 \end{center}
 \caption{\label{fig3} \small
LDA+SO band structure of \sriro , projected
on the \jeff12 (left panel), \j32\ $|m_j|$=$3/2$ (middle), and \j32\ $|m_j|$=$1/2$ (right panel)
spin-orbitals. }
\end{figure}

\begin{figure}
 \begin{center}
  \includegraphics[width=\linewidth,keepaspectratio]{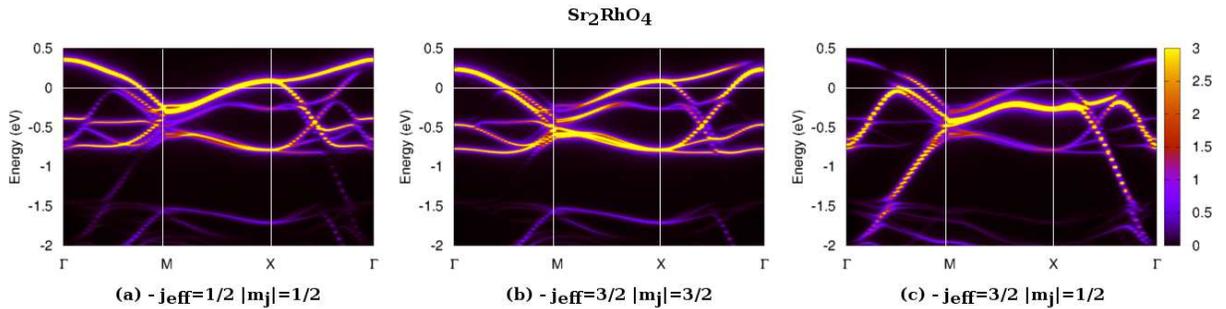}
 \end{center}
 \caption{\label{fig4} \small
LDA+SO band structure of \srrho , projected
on the \jeff12 (left panel), \j32\ $|m_j|$=$3/2$ (middle), and \j32\ $|m_j|$=$1/2$ (right panel)
spin-orbitals. }
\end{figure}

Fig.~\ref{fig3} and~\ref{fig4} depict the projection of these Wannier functions on the LDA+SO band structure. The similarities between Fig~\ref{fig3} and Fig~\ref{fig2}-(c) are numerous, thus confirming our previous band character analysis. Tab.~\ref{tab1} gives the decompostion of these local Wannier functions on the t$_{2g}$ manifold and their respective occupation. 

To obtain deeper insights into the nature of these Wannier orbitals, Tab.~\ref{tab0} gives the coefficients of the local Wannier orbitals obtained from the LDA+SO band structure of "undistorted" \sriro\ using an energy window $[-3.5,0.8]$\,eV.
The results agree well with the standard $j_\textrm{eff}$ picture (cf. Eq.~\eqref{eq-jeff12} and \eqref{eq-jeff32}) in both modulus and phase. Discrepancies are mostly due to the elongation of the IrO$_6$ along the $c$-axis, which introduces an additional tetragonal field between the t$_{2g}$ states. This effect also explains the lifting of the degeneracy of the 
two \j32 ($m_j=\pm 1/2$ and $m_j=\pm 3/2$) states and implies
the reason why the \jeff12\ is slightly more than half-filled. 

\begin{table}
\small
\begin{center}
\begin{tabular}{|c|c c c|}
\hline
\hline
  & & & \\
  "undistorted" \sriro & $\displaystyle\left|\frac{1}{2},\pm\frac{1}{2}\right\rangle$ & 
  $\displaystyle\left|\frac{3}{2},\pm\frac{1}{2}\right\rangle$ & $\displaystyle\left|\frac{3}{2},\pm\frac{3}{2}\right\rangle$ \\
  & & & \\
\hline
  $|$\dxy $\uparrow\downarrow\rangle$ & $\pm$ 0.6605 & $+$0.7508 & 0 \\
  $|$\dxz $\uparrow\downarrow\rangle$ & $\pm$ 0.5309~$i$ & $-$0.4670$i$ & $-$0.7071~$i$ \\
  $|$\dyz $\uparrow\downarrow\rangle$ & $+$ 0.5309 & $\mp$0.4670 & $\mp$0.7071 \\
\hline
  occupation (LDA+SO) & 1.20 & 1.92 & 1.86 \\
\hline
\hline
\end{tabular}
\caption{\label{tab0} \small 
Coefficients and occupation of the $j_\textrm{eff}$ Wannier orbitals in "undistorted" \sriro .
The discrepancy between these coefficients and those given in Eq.~\eqref{eq-jeff12} and \eqref{eq-jeff32} are due to the small elongation of the octahedra along the $c$-axis. 
}
\end{center}
\end{table}

Because of the hydridization between the $d_{xy}$ and $d_{x^2-y^2}$ orbitals in the distorted structures, we had to define in practice ``effective \jeff12\ and \j32\ $|m_j|=1/2$ states'', which remain close to the atomic $j_\textrm{eff}$ picture but take into account a small amount of $d_{x^2-y^2}$ character (cf. Tab.~\ref{tab1}). 
The coefficients have been calculated such that the density matrix
of the local atomic problem is closest possible to diagonal form.\footnote{With the obtained coefficients, the off-diagonal terms remaining in the local Green’s functions between the \jeff12\ and \j32\ $|m_j|=1/2$ are smaller than $0.05$. In practice, the coefficients were chosen real. This can be done in the local problem since only density-density terms were kept for the interaction terms and off-diagonal terms of the density matrix were neglected.} In addition to the hybridization, the construction of the "effective $j_\textrm{eff}$" takes also into account the tetragonal crystal field due to the elongation of the octahedra in each crystal structure: this explains the discrepancies with the standard coefficients given in Eq.~\eqref{eq-jeff12} and \eqref{eq-jeff32}.
We note that the coefficients obtained for the \jeff12\ state of \sriro\ are equivalent to those obtained in the AF phase in Ref.~\cite{JinPRB80-09}. 

Finally, comparing the occupation of the orbitals in Tab.~\ref{tab0} and \ref{tab1} highlights again the role of the hybridisation between the $d_{xy}$ and $d_{x^2-y^2}$ orbitals which pushes the band \j32\ $|m_j|=1/2$ further below the Fermi level close to $\Gamma$ : as a result, the four bands that cross the Fermi level are formed only by the  \jeff12\ and \j32\ $|m_j|$=$3/2$ orbitals and the \jeff12\ tend to be close to half-filling. Similar conclusions were drawn for the AF phase within a Variational Cluster Approximation (VCA) approach in Ref.~\cite{WatanabePRL105-10}. 
Similar conclusions hold for \srrho. 

\begin{table}
\small
\begin{center}
\begin{tabular}{|c|c c c||c c c|}
\hline
\hline
   & \multicolumn{3}{|c|}{\sriro} & \multicolumn{3}{|c|}{\srrho} \\
\hline
   & & & & & & \\
  Wannier orbitals &  $\displaystyle\left|\frac{1}{2},\pm\frac{1}{2}\right\rangle$ & 
  $\displaystyle\left|\frac{3}{2},\pm\frac{1}{2}\right\rangle$ & $\displaystyle\left|\frac{3}{2},\pm\frac{3}{2}\right\rangle$ &  $\displaystyle\left|\frac{1}{2},\pm\frac{1}{2}\right\rangle$ & 
  $\displaystyle\left|\frac{3}{2},\pm\frac{1}{2}\right\rangle$ & $\displaystyle\left|\frac{3}{2},\pm\frac{3}{2}\right\rangle$ \\
  & & & & & & \\ 
\hline
  $|$\dx2y2 $\uparrow\downarrow\rangle$ & 0.0388 & 0.0766 & 0 & 0.0100 & 0.0302 & 0 \\
  $|$\dxy $\uparrow\downarrow\rangle$ & 0.4499 & 0.8889 & 0 & 0.3153 & 0.9485 & 0 \\
  $|$\dxz $\uparrow\downarrow\rangle$ & 0.6309 & 0.3193 & 0.7071 & 0.6710 & 0.2231 & 0.7071 \\
  $|$\dyz $\uparrow\downarrow\rangle$ & 0.6309 & 0.3193 & 0.7071 &  0.6710 & 0.2231 & 0.7071 \\
\hline
  occupation (LDA+SO) & 1.16 & 1.98 & 1.84 & 1.42 & 1.96 & 1.64 \\
  charge (LDA+SO+DMFT) & 1.02 & 2.00 & 1.98 & 1.26 & 1.98 & 1.76 \\ 
\hline
\hline
\end{tabular}
\caption{\label{tab1} \small
Modulus of the coefficients of the $j_\textrm{eff}$ Wannier orbitals in \sriro\ and \srrho .
The occupation within LDA+SO and within LDA+SO+DMFT of each atomic Wannier orbitals is also provided, showing how electronic correlations enhance the spin-orbital polarization.}
\end{center}
\end{table}

\subsection{Effective Hubbard Interactions from cRPA}

After defining the $j_\textrm{eff}$ Wannier orbitals, we evaluate the local Coulomb interaction in the effective atomic problem within cRPA \cite{AryasetiawanPRB70-04,VaugierPRB86-12}, as explained in section~\ref{part-Ueff}.
For reasons of computational resources, the cRPA calculations were performed
in the case without distortions (without the rotations of the octahedra, 
hence considering only one formula-unit in a unit-cell) and without SOC. 
To mimick the effect of the distortions, the \eg\ states are shifted up 
to their energetic position in the presence of distortions.
We find $\mathcal{U}=2.54$~eV and $\mathcal{J}=0.23$~eV for \sriro\ and $\mathcal{U}=1.94$~eV and $\mathcal{J}=0.23$~eV for \srrho . These parameters lead to the following local interaction matrices for \sriro : 
\begin{equation}
 U^{m_j m_{j'}}_{jj'}=\left(\begin{array}{c|cc}
               0 & 2.08 & 2.21 \\ 
               \hline
               2.08 & 0 & 1.93 \\
               2.21 & 1.93 & 0 \\
         \end{array}\right)
\quad 
 U^{m_j\overline{m}_{j'}}_{jj'}=\left(\begin{array}{c|cc}
               2.25 & 1.98 & 1.90 \\ 
               \hline
               1.98 & 2.38 & 2.03 \\
               1.90 & 2.03 & 2.31 \\
         \end{array}\right)
\end{equation}
and for \srrho :
\begin{equation}
 U^{m_j m_{j'}}_{jj'}=\left(\begin{array}{c|cc}
               0 & 1.48 & 1.66 \\ 
               \hline
               1.48 & 0 & 1.29 \\
               1.66 & 1.29 & 0 \\
         \end{array}\right)
\quad 
 U^{m_j\overline{m}_{j'}}_{jj'}=\left(\begin{array}{c|cc}
               1.67 & 1.32 & 1.27 \\ 
               \hline
               1.32 & 1.86 & 1.46 \\
               1.27 & 1.46 & 1.71 \\
         \end{array}\right)
\end{equation}
where the values are in eV and the ordering of the  $|j,|m_j|\rangle$ orbitals is: $|1/2,1/2\rangle$,$|3/2,1/2\rangle$, $|3/2,3/2\rangle$ and $\overline{m}_j$ denotes $-m_j$. We remind the reader that 
$ U^{m_j m_{j'}}_{jj'}=U^{\overline{m}_j\overline{m}_{j'}}_{jj'}$ and $U^{m_j\overline{m}_{j'}}_{jj'}=U^{\overline{m}_jm_{j'}}_{jj'}$. Since we have used "effective $j_\textrm{eff}$" Wannier orbitals instead of the standard definition given in Eq.~\eqref{eq-jeff12} and \eqref{eq-jeff32}, some discrepancies with the formulae given in Eq.~\eqref{eq-Ueff} and in \cite{AhnKunesJPCM27-15} can be observed. 
Contrary to common belief, the Hubbard interactions are smaller in the $4d$-TMO than in its $5d$-counterpart. This might seem counterintuitive at first sight,
since the $5d$-orbitals are more extended than the $4d$ ones, but finds its
explanation in more efficient screening in the $4d$ material:
As shown in Fig.~\ref{fig2}-(d) and (e), the hybridization between the Rh-$4d$ states and the O-$2p$ is weaker in \srrho\ than in \sriro. 
Correspondingly, the energetic position of the O-$2p$ bands is closer to
the Fermi level by about $1$~eV, and as a result, the Coulomb interactions are screened more efficiently in \srrho\ than in \sriro, explaining the observed
trend.

\subsection{Correlated electronic structure of \sriro\ and \srrho}\label{part-spinpol}

DFT+DMFT calculations following the procedure described in 
section~\ref{part-ldadmft} indeed find an insulating solution for \sriro\ 
and a correlated metal for \srrho \cite{MartinsPRL107-11}, in agreement 
with experiment.
The difference in these metallic versus insulating
nature of \srrho\ and \sriro\ can be traced back to the different spin-orbital polarization in the 
three $j_\textrm{eff}$ orbitals, which is enhanced by Coulomb correlations.

The occupations of the $j_\textrm{eff}$ Wannier orbitals within LDA+SO and LDA+SO+DMFT are provided in Tab.~\ref{tab1}. In \sriro , one detects a considerable spin-orbital polarisation already at the LDA+SO level: the four \j32\ states are almost filled with $n_{3/2,|1/2|}$=$1.98$ and $n_{3/2,|3/2|}$=$1.84$ while the \jeff12\ states thus slightly exceed half-filling with $n_{1/2}$=$1.16$ (as in the "ideal undistorted" case). Taking into account Coulomb correlations within DMFT opens a gap of about $0.26$\,eV \cite{MartinsPRL107-11} and enhances the spin-orbital polarisation, such as to fill the \j32\ states entirely, leading to a half-filled \jeff12 state. This is thus the celebrated "\jeff12 -picture" \cite{KimPRL101-08}, which comes out here as a result of the calculations, rather than being an input as in most model Hamiltonian calculations.

A different picture emerges for \srrho\ according to Tab.~\ref{tab1}: while the spin-orbital occupations display some polarisation at the LDA+SO level, the smaller SOC -- and thus the smaller effective splitting between the $j_\textrm{eff}$ bands -- leads to a picture where only the \j32 $|m_j|=1/2$ state is entirely filled while both \j32 $|m_j|=3/2$ and \jeff12 live at the
Fermi level. 
This spin-orbital polarization is enhanced by Coulomb correlations
-- just as in \sriro\ -- but this enhancement is not enough to fill both \j32 states entirely and obtain a half-filled \jeff12 state. 
The higher effective degeneracy together with the 
smaller value of $\mathcal{U}$ eventually leave \srrho\ metallic.

\subsection{Spectral properties of Sr$_2$RhO$_4$: theory vs. experiment}\label{part-sro}

We now turn to the calculated spectral function of the spin-orbital correlated metal \srrho\ that we analyse in comparison to experiment.

\begin{figure}[t]
 \begin{center}
  \includegraphics[scale=0.3,keepaspectratio]{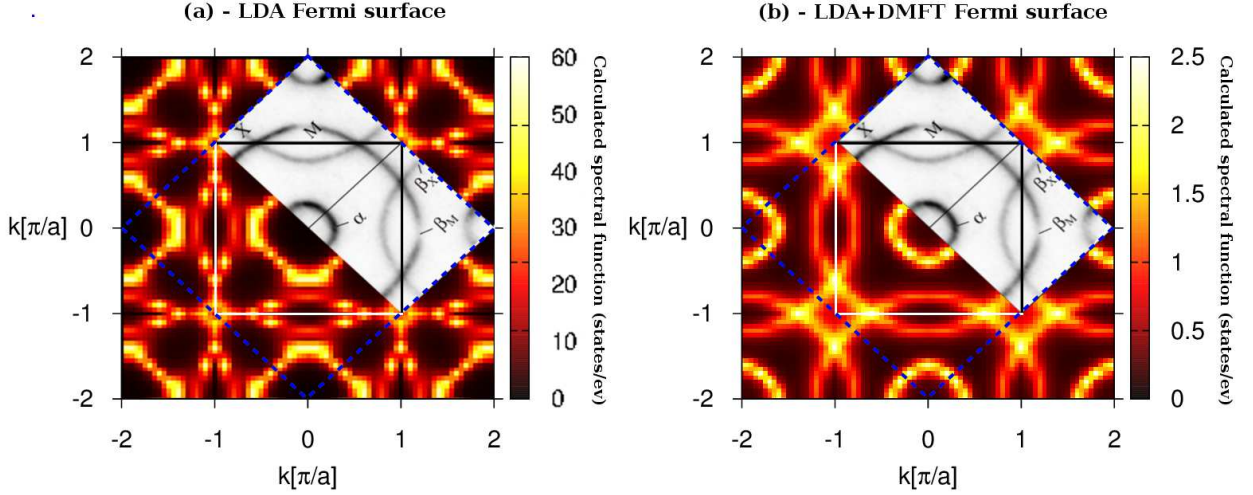}
 \end{center}
 \caption{\label{fig7} \small Calculated Fermi surface of \srrho\ in the $\mathbf{k}_z=0$ plane within LDA+SO (left panel) and LDA+SO+DMFT (right panel). Superimposed is the experimentally measured Fermi surface, from Ref.~\cite{BaumbergerPRL96-06}.}
\end{figure}

\begin{table}
\small
\begin{tabular}{|c|c c c|c c c|c c c|}
\hline
\hline
  &  \multicolumn{3}{|c|}{$\alpha$} & \multicolumn{3}{|c|}{$\beta_X$} & \multicolumn{3}{|c|}{$\beta_M$} \\
  &  LDA & DMFT &  Exp. &  LDA & DMFT &  Exp. &  LDA & DMFT &  Exp. \\
\hline
  FS volume $A$ (\% BZ) & 18.4 & 10.1 & 6.1(4) & 4.5 & 6.2 & 8.1(5) & 10.0 & 7.6 & 7.4(4) \\
  $\hbar \overline{v_F}$ (eV.\AA) & 1.252 & 0.645 & 0.41(4) & 1.260 & 0.674 & 0.55(6) & 1.260 & 0.674 & 0.61(6) \\
  $m^*$ ($m_e$) & 1.70 & 2.44 & 3.0(3) & 0.83 & 1.83 & 2.6(3) & 1.24 & 2.02 & 2.2(2) \\
\hline
\hline
\end{tabular}
\caption{\label{tab4} \small Comparison of the Fermi surface (FS) parameters evaluated within LDA+SO, within LDA+SO+DMFT and ARPES \cite{BaumbergerPRL96-06}. For each $\alpha$, $\beta_X$ and $\beta_M$ pocket, the FS volume $A$ is defined as a pourcentage of the two-dimensional-BZ volume (using the experimental lattice parameters ($a=5.45~\AA$)). The Fermi velocity $\hbar \overline{v_F}$ is obtained from the slope of the band dispersion at the Fermi level. The cyclotron mass $m^*/m_e$ is calculated using the same method as described in \cite{BaumbergerPRL96-06}:  $m^* \overline{v_F}=\hbar\sqrt{A/\pi}$.}
\end{table}

Fig.~\ref{fig7} depicts the Fermi surface of \srrho\ within LDA+SO (left panel) and LDA+SO+DMFT (right panel) in the $\mathbf{k}_z=0$ plane, to which we superimpose the experimental measurement from \cite{BaumbergerPRL96-06}. Tab.~\ref{tab4} gives more quantitative insight to ease the comparison between the different topologies. All three Fermi surfaces, the two theoretical ones and the experimental one, are {\it qualitatively} similar with three closed contours : a circular hole-like $\alpha$-pocket around $\Gamma$, a lens-shaped electron pocket $\beta_M$ and a square-shaped electron pockets $\beta_X$. These two structures merge in the undistorted tetragonal zone (dashed blue line in Fig.~\ref{fig7}) to a large electron-like pocket $\beta$.

Comparing Fig.~\ref{fig7}-(a) and (b) highlights the key role of electronic correlations : they decrease the radius of the $\alpha$ pocket from $0.26-0.29$~\AA$^{-1}$ to $0.21$~\AA$^{-1}$ and decrease the radius of the large $\beta$ pocket from $0.69-0.72$~\AA$^{-1}$ to $0.67-0.70$~\AA$^{-1}$, thus enlarging the $\beta_M$ and $\beta_X$ pockets such that their volume are well-reproduced within LDA+SO+DMFT (cf. Tab~\ref{tab4}). As a result, the agreement between LDA+SO+DMFT data and the experimental measurements becomes even quantitatively excellent. 

\begin{figure}[t]
 \begin{center}
  \includegraphics[scale=0.6,keepaspectratio]{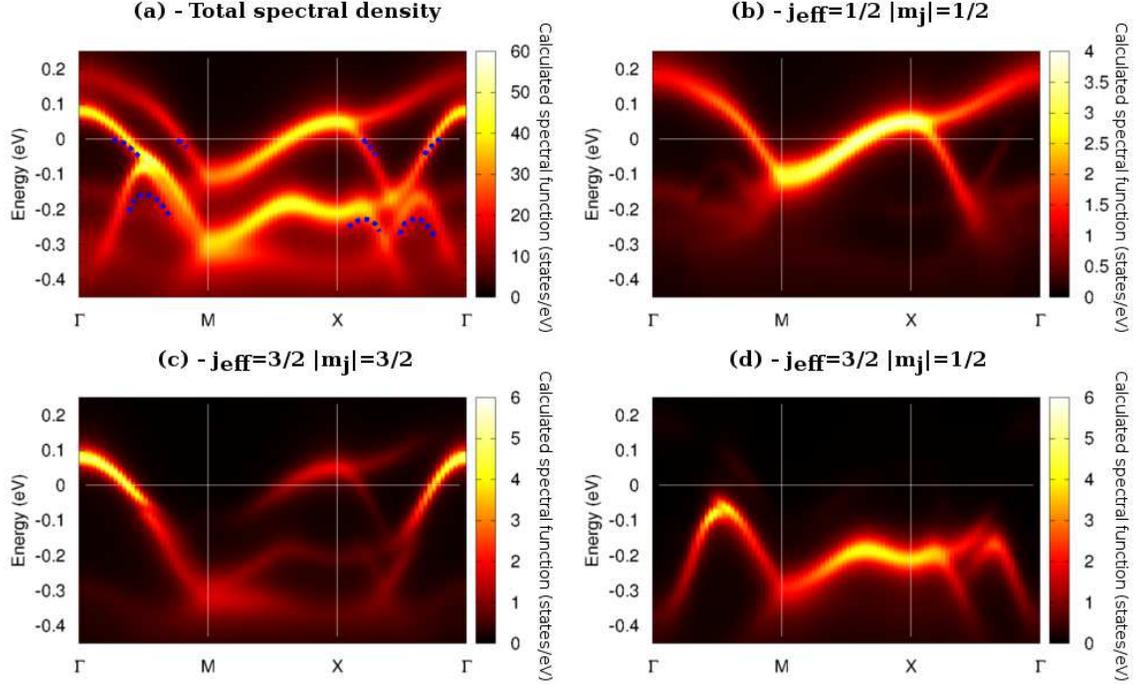}
 \end{center}
 \caption{\label{fig-srrhoAkw} \small Calculated momentum-resolved spectral function of \srrho\ within LDA+DMFT (a) and its orbital-resolved versions for the \jeff12 states (b), the \j32 $|m_j|=3/2 $ (c) and the completely-filled \j32 $|m_j|=1/2 $ (d). The dashed blue line on panel (a) are the reproduction of the ARPES structure from \cite{PerryNJP9-06}}
\end{figure}

To go further in the analysis, Fig.~\ref{fig-srrhoAkw} depicts the momentum-resolved spectral function, as well as its orbital-resolved version. The completely filled \j32 $|m_j|=1/2 $ state is visible (panel d), as well as the partially filled character of the \j32 $|m_j|=3/2 $ (panel c) and \jeff12 states (panel b). A detailed comparison with angle-resolved photoemission data from \cite{PerryNJP9-06} (blue dashed line on the figure) shows that the band dispersion around the Fermi level is well-reproduced, while some discrepancies are observed for the structures experimentally observed along $\Gamma-X$ and $\Gamma-M$ at lower energy. These features, reminiscent of the \j32\ $|m_j|=1/2$ bands, are indeed located about $0.05$\,eV higher in energy in our calculated spectral function. 

From Fig.~\ref{fig-srrhoAkw}-(b) and (c), one observes that the Fermi level is crossed by the renormalized \j32\ $|m_j|=3/2$ band at $0.20$~\AA$^{-1}$ along $\Gamma-X$ and at $0.21$~\AA$^{-1}$ along $\Gamma-M$, while the renormalized \jeff12\ band is responsible for all other crossings. 
This allows to label the hole-like $\alpha$-pocket as being of \j32\ $|m_j|=3/2$ type, whereas the two other pockets $\beta_M$ and $\beta_X$ are mostly of type \jeff12 . Using the quasiparticle weight of each state ($Z_{1/2}=0.535$ and $Z_{3/2,|3/2|}=0.675$), we evaluate the Fermi velocity at each crossing along the path $[\Gamma M X \Gamma]$: we find a huge variation of the values depending on $\mathbf{k}$ and give in Tab.~\ref{tab4} their mean value over the Brillouin zone.
Finally, using the same method as described in \cite{BaumbergerPRL96-06}, we evaluate the cyclotron mass $m^*/m_e$ based on the approximate formula used there:  $m^* \overline{v_F}=\hbar\sqrt{A/\pi}$. The DMFT results shown in Tab.~\ref{tab4} 
show a substantial improvement over DFT when compared to experiments.

\section{The effective orbital degeneracy as a key quantity determining the correlation strength}

In section~\ref{part-spinpol}, we have identified the spin-orbital 
polarization as a key factor 
to explain the different behavior of \srrho\ and \sriro .

In \sriro, 
Coulomb correlations enhance the spin-orbital polarisation, such as to fill the \j32\ states entirely, leading to a half-filled \jeff12 one-band picture, while in \srrho\
the final situation is an effective two-orbital system containing three
electrons.
This situation is akin to correlation-induced enhancements
of orbital polarisation also observed in other transition metal
oxides. In the distorted 3d$^1$ perovskites LaTiO$_3$ and YTiO$_3$,
for example, it was argued \cite{PavariniPRL92-04} that the interplay
of structural distortions and Coulomb correlations leads to
a suppression of orbital fluctuations in the t$_{2g}$-manifold,
favoring a particular orbital composition selected by crystal
and ligand field effects. At the LDA level, 0.45 [0.88] electrons
are found in this particular orbital in LaTiO$_3$ [YTiO$_3$],
while Coulomb correlations as described by LDA+DMFT lead to
an occupation of 0.88 [0.96] electrons.

In these systems, this reduction of effective orbital
degeneracy was shown to be key to their insulating
nature since the critical interaction strength needed to localise
the single electron is thus effectively determined by the one
of a single-orbital system, instead of the one of a three-fold
degenerate t$_{2g}$-manifold. Within DMFT, the critical Hubbard
interaction scales with the square-root of the orbital degeneracy $N$
for the lower critical interaction of the phase coexistence region
of the first order Mott transition, while the upper critical
interaction varies with $N$ \cite{Florens}.

Localising electrons in a single-orbital system therefore needs
a critical interaction roughly smaller by a factor of $3$ as
compared to the degenerate case. This was demonstrated to be
crucial for the difference in behaviors in the series of d$^1$
compounds SrVO$_3$, CaVO$_3$, LaTiO$_3$, YTiO$_3$, where the
former are three-fold degenerate metallic systems, whereas the
latter realise the single-orbital Mott state.

The situation in the iridates is analogous with the purely
formal difference that one is dealing with a one-hole
situation instead of one electron. Furthermore, the strong
spin-orbit interaction is instrumental for the suppression of 
the degeneracy, which is the net result of structural distortions,
spin-orbit coupling and Coulomb correlations.

This discussion highlights an important aspect of the physics
of transition metal oxides, often neglected when considering
band filling and interaction strength only: the {\it effective
orbital degeneracy} is a crucial tuning parameter for electronic
behavior, suggesting that the popular picture 
distinguishing filling-controlled and bandwidth-controlled
Mott transitions \cite{ImadaRMP70-98}
should be complemented by a ``third axis'' and the notion of
degeneracy-controlled Mott behavior.

Crystal and ligand fields together with spin-orbit coupling
and the Coulomb correlations themselves are the driving forces 
for establishing a given effective degeneracy.
At the level of the calculations, this effective degeneracy
is both an outcome of the calculation and a determining factor
of the properties of the given compound.

\section{Conclusions and Perspectives}

The common belief about electronic Coulomb correlations being less
important in 4d and 5d compounds as compared to 3d transition metal
oxides, was overruled by insights into the role of spin-orbit
coupling for the insulating behavior of iridates \cite{KimPRL101-08} and for 
the Fermi surface topology of \srrho\ \cite{LiuPRL101-08,HaverkortPRL101-08}.

Here, we have reviewed recent work on a first principles many-body description
of such effects within a dynamical mean-field framework.
We have highlighted the notion of the {\it effective degeneracy}
of the system as a crucial parameter determining the physical properties
of a system. The effective degeneracy is the result of a complex 
interplay of structural distortions spin-orbit coupling and Coulomb
correlations. We have stressed the analogy of the \jeff12 Mott
insulating picture for \sriro\ with the insulating nature of
LaTiO$_3$ and YTiO$_3$ in the ``degeneracy-controlled
Mott transition'' series of d$^1$ perovskites (SrVO$_3$,
CaVO$_3$, LaTiO$_3$, YTiO$_3$).

In \sriro\ and \srrho\ the difference in degeneracy is itself a 
consequence of
the quantitative aspects of the physics of these two compounds:
all three decisive elements -- structural distortions, spin-orbit
coupling and Hubbard interaction -- are smaller in \srrho\ than
in \sriro\ and this quantitative difference in the electronic
parameters translates into a qualitative difference of the
resulting properties.

We have analysed in detail the spectral properties of \srrho ,
a spin-orbit correlated 4d metal where the effective degeneracy is reduced
by spin-orbit coupling and correlations but not to the point
such as to induce a \jeff12 Mott insulator.
The calculated spectral properties and Fermi surface are in 
excellent agreement with experimental data.
A detailed analysis of the spectral properties of \sriro\ 
is left for future work.

\section{Acknowledgments}

This work was supported by the ERC Consolidator Grant
CORRELMAT (grant 617196), the French ANR under project
IRIDATES, and IDRIS/GENCI under project t20169313. M.A. is supported by a START program of the Austrian Science Fund (FWF), grant number Y746.

\section*{Appendix A: Generalized partial \emph{$\Theta$-projectors} and spectral function}

In order to calculate quantities for a given atom $\alpha$ and a particular orbital (spin) character $j$ ($m_j$) -- such as the spectral functions $A_{j}^{m_j\alpha}(\mathbf{k},\omega)$ -- , a set of partial projectors called ``\emph{$\Theta$-projectors}'' was built. Contrary to the previously introduced Wannier projectors $P^{\alpha,m_j}_{j,\nu}(\mathbf{k})$, their definition is not restricted to the correlated orbitals only. The formalism of these  partial projectors was initially introduced in \cite{AichhornPRB80-09} and was extended 
to the case where spin is not a good quantum number anymore in
\cite{MartinsPRL107-11}.\\

Inside the muffin-tin sphere associated to an atom $\alpha$, one can write the spin-$\sigma$ contribution of the eigenstate $\psi_{\mathbf{k}\nu}(\mathbf{r})$ as:
\begin{eqnarray} \label{eq-Theta1}
\phi_{\mathbf{k}\nu}^{\sigma}(\mathbf{r}) & = & \sum_{\ell=0}^{\ell_{max}}\sum_{m=-\ell}^{+\ell} \displaystyle \Big[ A^{\nu\alpha}_{\ell m}(\mathbf{k},\sigma)~ u_{\ell m,1}^{\alpha,\sigma}(\mathbf{r}^\alpha) + B^{\nu\alpha}_{\ell m}(\mathbf{k},\sigma)~u_{\ell m,2}^{\alpha,\sigma}(\mathbf{r}^\alpha) + C^{\nu\alpha}_{\ell m}(\mathbf{k},\sigma)~u_{\ell m,3}^{\alpha,\sigma}(\mathbf{r}^\alpha) \Big]\nonumber\\
\end{eqnarray}

where the basis $\{u_{\ell m,i}^{\alpha,\sigma}\}_{i=\{1,2,3\}}$ is not orthonormalized as already mentioned in~\cite{AichhornPRB80-09}. That is why, to make the calculations easier, one introduces an orthonormal basis set $\{v^{\alpha,\sigma}_{\ell m,j}\}_{\{j=1,2,3\}}$ for each atomic orbital $(\ell,m)$. These orbitals are defined from the initial basis $\{u_{\ell m,i}^{\alpha,\sigma}\}_{i=\{1,2,3\}}$ as follows:
\begin{equation}
 \forall i \quad u_{\ell m,i}^{\alpha,\sigma}(\mathbf{r}^\alpha)=\sum_{j=1}^3 c_{ij}v^{\alpha,\sigma}_{\ell m,j} \quad \textrm{with} \quad \mathbf{C}=\left(\begin{array}{ccc}
1 & 0 & \langle u_{\ell m,1}^{\alpha,\sigma}|u_{\ell m,2}^{\alpha,\sigma}\rangle \\
0 & \langle u_{\ell m,2}^{\alpha,\sigma} |u_{\ell m,2}^{\alpha,\sigma} \rangle  & \langle u_{\ell m,2}^{\alpha,\sigma} |u_{\ell m,3}^{\alpha,\sigma}\rangle \\
\langle u_{\ell m,3}^{\alpha,\sigma}|u_{\ell m,1}^{\alpha,\sigma}\rangle  & \langle u_{\ell m,3}^{\alpha,\sigma}|u_{\ell m,2}^{\alpha,\sigma}\rangle  & 1 \end{array}\right)^{\frac{1}{2}}.
\end{equation}

We can then rewrite \eqref{eq-Theta1} as:
\begin{equation}
\psi_{\mathbf{k}\nu}^{\sigma}(\mathbf{r})= \sum_{\ell=0}^{\ell_{max}}\sum_{m=-\ell}^{+\ell}~\sum_{i=1}^3 \Theta_{\ell m\nu,i}^{\alpha,\sigma}(\mathbf{k}) v^{\alpha,\sigma}_{\ell m,i}(\mathbf{r}^\alpha).
\end{equation}
The matrix elements $\Theta_{\ell m\nu,i}^{\alpha,\sigma}(\mathbf{k})$ are the ``\emph{$\Theta$-projectors}'', which are thus defined by:
\begin{equation} \label{eq-Theta2}
\Theta_{\ell m\nu,i}^{\alpha,\sigma}(\mathbf{k}) = \langle v^{\alpha,\sigma}_{\ell m,i} | \phi_{\mathbf{k}\nu}^{\sigma} \rangle = A^{\nu\alpha}_{\ell m}(\mathbf{k},\sigma)c_{1i} + B^{\nu\alpha}_{\ell m}(\mathbf{k},\sigma)c_{2i} + C^{\nu\alpha}_{\ell m}(\mathbf{k},\sigma)c_{3i}.
\end{equation}

Contrary to the implementation of \cite{AichhornPRB80-09}, there is now a couple of $\Theta$-projectors associated to each band index $\nu$, 
$\Theta_{\ell m\nu,i}^{\alpha,\sigma}(\mathbf{k})$ with $\sigma=\uparrow,\downarrow$, since spin is not a good quantum number anymore.

We have introduced here the $\Theta$-projectors in the complex spherical harmonics basis. As for the Wannier projectors, it is of course possible to get the $\Theta$-projectors in any desired $j,m_j$ basis:
\begin{equation}
\Theta_{j\nu,i}^{\alpha,m_j}(\mathbf{k}) =  \sum_{m,\sigma} \mathcal{S}_{j,\ell m}^{m_j,\sigma}\Theta_{\ell m\nu,i}^{\alpha,\sigma}(\mathbf{k})
\end{equation} 

Finally, the spectral function $A(\mathbf{k},\omega)$ which is defined by:
\begin{equation}
A(\mathbf{k},\omega)=-\frac{1}{\pi}\mathfrak{Im}\left[G(\mathbf{k},\omega)\right].
\end{equation}
is obtained for a given atom $\alpha$ with orbital character $(j,m_j)$ through the following formula:
\begin{equation}
 A_{j}^{\alpha,m_j}(\mathbf{k},\omega) = \displaystyle -\frac{1}{\pi}\mathfrak{Im}\left[\sum_{\nu\nu'} \sum_{i=1}^3 \mathbf{\Theta}_{j\nu,i}^{\alpha,m_j}(\mathbf{k}) \mathbf{G}_{\nu\nu'}(\mathbf{k},\omega+i0^+) \left[\mathbf{\Theta}_{j\nu',i}^{\alpha,m_j}(\mathbf{k})\right]^*\right]
\end{equation}
where the band indices $\nu$, $\nu'$ run over \textit{both} spin and orbital quantum number.

\bibliographystyle{elsarticle-num}
\bibliography{./highlight_all12}

\end{document}